\documentstyle[referee,epsfig,graphicx,color]{mn2e_myversion}

\newif\ifAMStwofonts
\newcommand{\be}{\begin{equation}}
\newcommand{\ee}{\end{equation}}
\newcommand{\bea}{\begin{eqnarray}}
\newcommand{\eea}{\end{eqnarray}}

\begin{document}

\title{\centerline {Spiral Structure in Scale-free, Thin Discs I: Rigid Rotation}}

\author[ R.N. Henriksen]
{
R.N. Henriksen$^1$\thanks{henriksn@astro,queensu.ca}}
\centerline{R.N. Henriksen}
\centerline{Queen's University, Kingston, Ontario, K7L 3N6,Canada}

\date{\today}
\pagerange{\pageref{firstpage}--\pageref{lastpage}}
\pubyear{2011}

\maketitle

\label{firstpage}

\begin{abstract}
  
In this paper we suggest the existence in the central regions of spiral galaxies of  collisionless, scale-free, rigidly rotating, self-gravitating discs with spiral symmetry.
Such  discs must be truncated at a finite radius , and they must be stabilized and rendered self-similar by a suitable halo. The halo and the rotating disc share the self-similar class and must form together to arrive at the suggested state.  We make comparisons with the well-known  rigidly rotating, Kalnajs discs; one of which is axi-symmetric and finite while the other is infinite and decomposed into spiral modes. We find the self-consistent, self-similar, distribution functions in one and two dimensions in a rigidly rotating, collisionless system. In the case of two dimensions we deduce the self-consistency condition for discrete spiral arms.
 We  give an estimate of the disturbance created in the halo by the presence of the disc, and argue that the halo itself should be close to self-similarity. A very weak cusp in the halo may be necessary. The necessary spatial coincidence of the halo results in a kind of disc-halo `conspiracy'. Finally the disc equations are formulated in `spiral' coordinates, and the passage to an approximately discrete `line spiral' is given as an example. Although in two dimensions the collisionless particles enter and leave the arms in non-linear epicycles, they move approximately parallel to the arms in the line spiral limit. The spiral pattern is however in rigid rotation. Aperiodic spiral arms are suggested wherein discontinuities may be coarse-grained to appear as collisionless shocks.   
       
     
\end{abstract} 
\title{Collisionless Thin Discs }

\setlength{\baselineskip}{13pt}
\section{Introduction}
\label{sec:intro}

Axially symmetric, self-similar, thin, self-gravitating discs have been well studied; particularly so in the works of Evans (1994),and of Evans and Read (1998a), which developed earlier work by Zang (1976). Such studies have revealed the surface-density/potential pairs and  have deduced  self-consistent surface distribution functions. Evans and Read (1998b) have investigated the stability of such discs by expanding the modes in logarithmic spirals.

The conclusions regarding the stability of strictly scale-free discs were left somewhat uncertain until a summary paper by Goodman and Evans (1999) in which scale-free gaseous discs were examined. It was argued there that without boundaries, the normal modes of the scale-free disc were not properly defined. This was consistent with Evans and Read (1998b)who imposed a `cutout' centre to study the normal modes. These modes tended to be neutral for bi-symmetric modes relative to the more rapidly growing one-armed or axially symmetric modes.

The object of such time-dependent perturbation analyses is naturally to clarify the origin of the spiral structure of late type galaxies. Rather than attempt this discussion in this series of papers, we will instead bypass this question of origin by constructing non-linear, non-axially symmetric, stationary, uniformly rotating, spirals. In general non-linear stability requires a suitable stabilizing halo for equilibrium. The halo remains passive in our study and is thus artificial, but the existence of rigidly rotating structures in the centres of galaxies may nevertheless render its properties of interest.  

The stability of the rigidly rotating, entirely self-consistent Kalnajs disc was studied originally in Kalnajs (1972) and as part of a more general study by Vauterin and Dejonghe (1995). This latter work is very close to the present one, including the assumption that the disc is not isolated but is rather coupled to an indefinite passive halo, be it either of dark matter or of a bulge of stars or both. This halo is such as to make the disc potential harmonic regardless of the disc distribution function. The interesting result of this previous work for our purposes is that with the Kalnajs DF (as defined below) the uniformly rotating disc is found to be linearly  stable in a sufficiently strong (relative to a maximum disc) constant density halo. No spiral structure develops. However they find this to be peculiar to the Kalnajs disc and NOT characteristic of uniformly rotating discs in general. In fact it is readily shown that the non-linear, non-axially symmetric distortions of a Kalnajs disc depend critically on both the central and edge cut-offs for equilibrium. 

Another work that uses surprisingly similar techniques is \cite{SS96}. These authors also consider scale-free discs, but they replace the the collisionless matter by a barotropic, two-dimensional fluid. Their scaling constraints are necessarily somewhat different because of this, but nevertheless their $\beta$ can be identified with our $\alpha-\delta$ (not to be indentified with our ad hoc use of $\beta$ in section 3) and our $\xi$ is their $\phi$.
They also discuss (redone independently in this work!) the scaling of the Poisson integral as do we in the section on the Kalnajs disc. The self-similar case that they identify with $\beta=1/4$ is our class $a=5/4$, while the special case $\beta=0$ is our case $a=1$. Their weakly non axially-symmetric limit corresponds to our limit of small winding angle $\epsilon/\delta$, which we do not elaborate in this work.

In the present work  we treat both the non-linear spiral arms and  the background axi-symmetric disc as self-similar (assuming the stabilizing halo) and collisionless. Thus we are taking the stars and possibly molecular clouds to be the dominant material component both of the disc and of the spiral arms. The resulting discs resemble locally  infinite discs, but in fact they must be globally finite and stabilized as alluded to above and in more detail, below . Embedded in a real galaxy they are probably cut-off in a ring-like structure before the corotation resonance. 

We have no novel answer to the `anti-spiral' objection (Lynden-Bell and Ostriker, 1967), other than to appeal to time-irreversible disturbances (e.g. `fly-by' encounters) and dissipation. In one of our non-axi-symmetric solutions we shall in fact find that, discontinuities which imply shock structure of some type are required. It should also be noted that collective effects in N-body systems (such as Landau damping) can lead to local perturbations which are soon amplified in their effects by Liapunov divergence. This leads to an effective irreversibility since virtually infinite accuracy in particle motion is soon required to time reverse  the system.

In this first paper we concentrate on the uniformly rotating disc, because it seems not to have been previously studied as a scale-free example (although the scale invariance of the Poisson integral was known; Kalnajs, 1971)  and because it allows a straight forward introduction of the basic ideas. We shall extend the treatment subsequently to a more general scale-free disc. Such a constantly rotating disc does not correspond to the outer structure of any late type galaxy, although it is relevant to the inner structure that may display spirals and bars. It can also be applied to some irregular galaxies.

Our treatment finds in most cases the surface-density/potential pair and the collisionless distribution function self-consistently by an `ab initio' self-similar analysis. In  section 2 including various sub-sections we give the usual (e.g. Carter and Henriksen, 1991; Henriksen, 1997,Le Delliou, Henriksen and MacMillan, 2011) self-similar formalism and deduce the general forms. Sections 3 and 4 deal with some axi-symmetric limits and make contact with the well known` Kalnajs disc' (Kalnajs, 1972; Binney and Tremaine, 1987, 2008). The non axi-symmetric Kalnajs discs whose potential follows from the decomposition of the Poisson integral into spiral modes, are also studied in these sections. Section 4 deals particularly with the halo structure required to permit  stable, self-similar centres of finite discs. In section 5 we construct various types of  stationary, spiral arms, including  discontinuous examples wherein shocks must appear. In section 6 a useful formulation of the problem in `spiral coordinates' is given and then is used to study isolated spiral arms in a stabilizing halo.  Finally we summarize and discuss our results.

\renewcommand{\textfraction}{0}
\renewcommand{\topfraction}{1} 
\renewcommand{\bottomfraction}{1}

\section{Self-Consistent, Self-similar Disc Equations}\label{sec:dynamics}
In this section we review the fundamental equations and deduce the self-similar forms for a rigidly rotating collisionless disc. We imagine it to be composed of stars and perhaps small molecular clouds. 

\subsection{Basic Equations }\label{sec:basic}

We work in cylindrical coordinates $\{r,\phi,z\}$ rather than in spherical coordinates because the algebra is simpler. However spherical coordinates are more appropriate if the disc is to be made compatible with external spherical mass distributions, and we use these in a brief discussion of the stabilizing halo below. 
The equations of the stationary thin disc are:

\bigskip
{\bf Collisionless Boltzmann Equation in the Disc }
\bigskip
 
We use consistently a reference frame that is rotating with uniform angular velocity $\Omega$, so that both the azimuthal velocity $v_\phi$ and the azimuthal angle $\phi$ are measured in this reference frame. We designate the two dimensional distribution function (DF) by $F(r,\phi,v_r,v_\phi)$ and the potential by $\Phi(r,\phi)$. We frequently use the notation $\partial_x$ for the operator $\partial/\partial_x$ where convenient. Other notations are standard. The equation is formally identical in spherical coordinates.
\be
v_r\partial_r F+\frac{v_\phi}{r}\partial_\phi F+(\frac{v_\phi^2}{r}+2\Omega v_\phi+\Omega^2 r-\partial_r\Phi)\partial_{v_r}F-(\frac{v_\phi v_r}{r}+2\Omega v_r+\frac{1}{r}\partial_\phi \Phi)\partial_{v_\phi}F=0.\label{eq:stacbe}
\ee
\bigskip
{\bf Laplace Equation Above the disc}
\bigskip
\be
\frac{1}{r}\partial_r(r\partial_r\Phi)+\frac{1}{r^2}\partial_\phi^2\Phi+\partial_z^2\Phi=0.\label{eq:Poisson}
\ee
\bigskip
{\bf Boundary Conditions for Laplace}
\bigskip

We designate the surface density of the disc by $\sigma(r,\phi)$. We require for self-consistency the relations

\be
(\partial\Phi)_{z=0}=2\pi G\sigma,\label{eq:boundary1}
\ee
 and 
\be
\sigma=\int~F~dv_rdv_\phi, \label{eq:boundary2}
\ee
where $F$ is a solution of equation (\ref{eq:stacbe}) using the potential $\Phi(r,\phi,0)\equiv \Phi_o(r,\phi)$. 

So far we have considered the disc as infinite in radius, but such a self-similar, uniformly rotating disc can not be in equilibrium\footnote{I am indebted to a referee for emphasizing this point}. Thus there is a radial boundary that must be considered, plus a surrounding halo that is required for equilibrium. We will discuss these matters in detail below, but it suffices here to say that for a finite disc a halo can be contrived that permits the central regions of the disc to be self-similar as described here. The required halo does not differ greatly from a uniform density spherical mass distribution, which is also of similarity class $\alpha=0$. 

This completes the set of basic equations. 

%


\subsection{Self-Similar Analysis}

We transform the basic equations to `scaling invariant variables' according to the usual analysis (see e.g \cite{LeDHM11}; \cite{H97}; \cite{HW95};\cite{CH91}). The transformation takes the form:\\

\bigskip
{\bf Independent Variables}
\bigskip

Here $\delta$ is the (reciprocal) spatial scaling factor and the mass scaling factor is related to it as $\mu=3\delta$, because $G$ is invariant and the time scaling factor $\alpha=0$. The latter is imposed by the invariance of $\Omega$ under the scaling. We retain the remaining scaling parameter $\delta$ explicitly below because we can sometimes simplify results in the coarse or fine grained limits as $\delta\rightarrow \infty$ or $\delta\rightarrow 0$ respectively. Apart from this possible use, we may set $\delta=1/r_o$ where $r_o$ is an arbitrary fiducial radius.
Thus the scale-invariant variables $R,Z,\xi$ are found from \cite{CH91}
\be
\delta r=e^{\delta R},~~~~~\xi=\phi+\epsilon R\equiv \phi+(\epsilon/\delta)\ln{\delta r}, ~~~~~ \delta_z z=Ze^{\delta R}.\label{eq:indepvars}
\ee

We observe that the non-axisymmetric self-similar variable is a logarithmic spiral in variables $\{r,\phi\}$ with  $rd\phi/dr=-\epsilon/\delta$ as the tangent of the `winding angle' equal to ninety degrees minus the pitch angle.  The extra scaling parameter $\delta_z$ can be useful in strict boundary-layer problems where it might be taken much smaller than $\delta$ in order to ignore the $z$ dependence in the potential, or much larger to accentuate it. We will normally take it to equal $\delta$ below.\\

\bigskip
{\bf Dependent Variables}
\bigskip

We transform the independent variables to functions of the scale invariants $\{R,\xi,Z\}$ according to the dimensions of each quantity while taking $R$ as the Lie scaling group parameter (see e.g Henriksen and Widrow (1975)) to write 
\bea
\Phi=e^{2\delta R}\Psi(R,\xi,Z),&  & F=e^{-\delta R}P(R,\xi,{\bf Y})\nonumber\\
\sigma=e^{\delta R}\Sigma(R,\xi),&  & {\bf v}= e^{\delta R}{\bf Y}.
\label{eq:depvars}
\eea
These follow our standard procedure where for example $F=\exp{(2\alpha-4\delta+\mu)}$ in terms of time, space and mass dimensions and scales. However $\alpha=0$ for fixed $\Omega$ and $\mu=3\delta-2\alpha$ to conserve Newton's constant under scaling, and so finally we obtain the specified $F=e^{-\delta R}P$.

It will be verified below in the section on Kalnajs discs that  the Poisson integral for the potential in the plane may be wholly expressed in terms of these variables.

We proceed by a direct substitution of these dependent and independent variables into our set of basic equations. We obtain the equations that the scale independent quantities must satisfy in the forms:\\

\bigskip
{\bf Self-Similar Boltzmann}
\bigskip

Our substitutions yield
\bea
Y_R\partial_R P&+&(\delta Y_\phi+\epsilon Y_R)\partial_\xi P+(\delta(Y_\phi^2-Y_R^2)+2\Omega Y_\phi+\Omega^2/\delta-\epsilon\partial_\xi\Psi-\partial_R\Psi-2\delta\Psi)\partial_{Y_R} P\nonumber\\
-(2\delta Y_{\phi}Y_R&+&2\Omega Y_R+\delta\partial_\xi\Psi)\partial_{Y_\phi} P=\delta Y_R P.\label{eq:sscbe}
\eea
 This equation is easily solved by characteristics (the self-similar particle trajectories), which will be carried out below.

\bigskip
{\bf Self-Similar Laplace}
\bigskip

We find that our substitutions yield for this equation 
\be
(1+(\frac{\epsilon}{\delta})^2)~\partial^2_\xi\Psi+\frac{4\epsilon}{\delta}\partial_\xi\Psi+\partial^2_Z\Psi+4\Psi=0.\label{eq:sslaplace}
\ee 

\bigskip
{\bf Self-Similar Boundary Conditions}
\bigskip

Using  $\delta_z=\delta$ the boundary conditions become
\be
2\pi G\Sigma=\delta(\partial_Z\Psi)_o,\label{eq:ssboundary1}
\ee
and
\be
\Sigma=\int~P~dY_RdY_\phi.\label{eq:ssboundary2}
\ee

It is important to realize that we impose the self-similarity ultimately by requiring the scale-invariant functions $P$,$\Psi$ and $\Sigma$ to be independent of the group parameter $R$. This has already been enforced explicitly in the self-similar Laplace equation.

\bigskip
{\bf Is the Self-Similarity Physical?}
\bigskip

A referee has properly questioned the physicality of an infinite self-similar disc on the grounds that the outer regions of such a disc always dominate the local gravitational field. This would lead locally to an infinite gravitational force directed outwards, in stark contradiction to the assumed equilibrium.

This correct argument implies that the self-similarity can only apply to a truncated disc of radius `R' at sufficiently small $r$, where one might not expect the disc radius to be physically able to destroy the self-similarity. In fact one finds that for most surface density profiles a suitable halo is still necessary to stabilize the disc and to impose the self-similar behaviour. A non axially symmetric disc is somewhat more suited to the self-similarity, but in general the halo and the finite disc must be self-similar together. We shall discuss these points in detail below.

In the next sub-section we study the scale-free Boltzmann equation in order to see how this symmetry restricts the normal Jeans' theorem.

\subsection{The Disc Scale-Free Distribution Function}

In this section we study the solution to equation (\ref{eq:sscbe}), with and without axial-symmetry. Proceeding generally, it is easily found that the characteristics of equation (\ref{eq:sscbe}) yield together
\bea
\frac{dP}{ds}&=& \delta Y_R P,\label{eq:Pchar}\\
\frac{d\cal E}{ds}&=& -2\delta Y_R \cal E,\label{eq:Echar}
\eea
where $ds$ is along the characteristic and $\cal E$ is the scale-free energy given by
\be
{\cal E}\equiv \frac{Y_R^2+Y_\phi^2}{2}+\Psi-\frac{\Omega^2}{2\delta^2}.\label{eq:ssE}
\ee 

These latter two equations may be combined to yield the general value for the scale-free distribution function in the form
\be
P=\frac{K}{|{\cal E}|^{1/2}},\label{eq:intssP}
\ee

where $K$ is strictly constant. It may depend on a characteristic constant if one can be found, but no such constant is available in the strictly self-similar equations.  

We may translate this result into non-scaled variables by using the transformations given in equations (\ref{eq:depvars}). In this way we find
\be
F=\frac{K}{|E|^{1/2}},\label{eq:intssF}
\ee
where the characteristic energy in the rotating frame is $E\equiv v^2/2+\Phi-\Omega^2r^2/2$. It is important to note that if we write explcitly the upper limit in energy as $E_o$, then in strict self-similarity this must be of the form $k\Omega^2 r^2$ where $k$ is a real number. This is due to a constant energy being incompatible with the $\alpha=0$ self-similarity class. 

It is instructive to relax the self-similarity  by allowing $P(R) $and $\Psi(R)$. This does not change the two equations (\ref{eq:Pchar}),(\ref{eq:Echar}) above (although it does change the Poisson equation) but one now obtains in addition that $Y_R=dR/ds$. This means that these latter two equations may be integrated directly to give ${\cal E}=Ee^{-2\delta R}$ and $P=Fe^{\delta R}$. We have in these expressions identified the two constants of integration to be the the characteristic energy $E$,  and the distribution function $F$, also constant on a characteristic. These identifications follow from the scalings of equations (\ref{eq:indepvars}) and (\ref{eq:depvars}). 

Consequently by substituting these integrals, equation (\ref{eq:intssP}) becomes 
\be
F=\frac{K(E)}{|E_o-E|^{1/2}}.\label{eq:intF}
\ee
Unlike the situation in the self-similar limit, one can take the energy $E_o$ to be constant. 

 Basically in this non-self-similar case, this says only that $F=F(E)$ as by the Jeans' theorem. However it does suggest a special case where $K(E)$ is constant. This proves to be useful in our discussion of the Kalnajs disc.

Should the solution we seek be axi-symmetric, the azimuthal velocity on the characteristic satisfies from equation (\ref{eq:sscbe})
\be
\frac{dY_\phi}{ds}=-2Y_R(\delta Y_\phi+\Omega),
\ee
which is not integrable in the self-similar case. However by dropping the self-similarity and using again $Y_R=dR/ds$ we obtain the integral
\be
\delta Y_\phi+\Omega=\delta^2j_ze^{-2\delta R},\label{eq:intj}
\ee
where once again we have used equations (\ref{eq:depvars}, \ref{eq:indepvars}) to identify the constant of integration as essentially the specific angular momentum $\delta^2 j_z$. This implies a two integral DF which we may take in the suggestive form 
\be
F=\frac{K(E,E/(\Omega j_z))}{|E_o-E|^{1/2}}.\label{eq:intaxiF}
\ee
By letting $E/(\Omega j_z)$ be universally constant we return to equation (\ref{eq:intF}), after which we may impose the self-similar form by taking $K(E)$ constant and setting $E_o=k\Omega^2r^2$. 

We recall that the energy in the rotating frame $E$ is related to the energy in the inertial frame $E_I$ by $E=E_I-\Omega j_z$, so that taking $E/(\Omega j_z)$ constant ensures that the ratio $E_I/\Omega j_z$ is also constant.
This latter constant is equal to unity in WKB density wave theory (see e.g. equation (6.86) of Binney and Tremaine 2008) and in fact also for the multi-poles of electromagnetic radiation (Jackson, 1999). The definitions are slightly different in each case, but a ratio of unity seems to reflect an essential relation between angular momentum `flux' and energy `flux' in a wave field. A uniformly rotating disc may be the `near zone' of such a wave field.
  
\section{The  `Kalnajs' Discs}

In this section we first compare the rigidly rotating Kalnajs disc to the (finite) self-similar disc. We note that although the surface density at small radius is not self-similar, both the potential and the self-consistent DF are.
Subsequently we discuss the Kalnajs \cite{K71} disc/potential decomposition into logarithmic spirals and show in what sense this approach is self-similar. These are not rotating discs in strict self-similarity since $\alpha\ne0$. In fact we find in subsequent work (noted below)  that the corresponding DF is isotropic in an inertial frame. 

\subsection{Rigidly Rotating Finite Disc}

For the usual Kalnajs disc it is convenient to drop the scale-free formulation. Formally with a fixed  angular velocity and a fixed radius both the temporal scaling $\alpha$ and the spatial scaling $\delta$ are zero. The only remaining self-similar `motion' is through the dependence on $\xi$, but this does not concern us in this section since we do not include spirality.  
   
This disc is defined by taking the flat limit of an axially symmetric MacLaurin spheroid of radius $a$ to find the surface density
\be
\sigma=\sigma_o\sqrt(1-\frac{r^2}{a^2}).\label{eq:Maclaurin}
\ee
For this reason it is also referred to as a `MacLaurin' disc. The remarkable property of this disc is that in the plane, and for $r\le a$, the potential has the same form as does the infinite self-similar disc namely
\be
\Phi=\frac{\Omega_c^2r^2}{2},\label{eq:McLpot}
\ee
where the circular angular velocity is given by
\be
\Omega_c^2=\frac{\pi^2G\sigma_o}{2a}.\label{eq:McLcirc}
\ee
This disc (Kalnajs, 1972) has been discussed recently in \cite{BT87} and \cite{BT08}  and an explicit form for the potential everywhere has also been given recently by \cite{Sch09}. In the latter work the potential found  in the plane for $r<a$ can be easily verified by using the Bessel function version of the density/potential pairs.

The effective potential in the rotating frame is 
\be 
\Psi_J\equiv \frac{\Omega_c^2-\Omega^2}{2}r^2,\label{eq:effpot}
\ee

It is easy to verify that for $r<a$ a consistent solution requires $\Omega_c>\Omega$ so that $\Phi_J> 0$. We note that because of symmety in $v_\phi$ in the  DF (\ref{eq:intF}), the average value of this quantity is zero. This means that the average circular velocity in the inertial frame is $\overline{v_\phi}=\Omega r$, giving $\Omega$ a physical interpretation. 

We can now close the loop by evaluating the surface density from the integral (\ref{eq:boundary2}) using the DF (\ref{eq:intF}) with general $E_o$ . This procedure yields with $\Phi_J>0$ and $E_o>0$ 
 \be
\sigma=4\pi K\sqrt{E_o-\Phi_J}=4\pi K\sqrt{(E_o-\frac{\Omega_c^2-\Omega^2}{2}r^2)},\label{eq:sigmaK}
\ee
and this must equal $\sigma_o\sqrt{1-r^2/a^2}$ for consistency. Insisting on the equality requires
\be
E_o=\frac{\Omega_c^2-\Omega^2}{2}a^2,\label{eq:KE_o}
\ee
which is always greater than $\Phi_J$ as is necessary; and the further condition on $K$ as
\be
4\pi K=\frac{\sigma_o}{\sqrt{(\frac{\Omega_c^2-\Omega^2}{2}a^2)}}\label{eq:KK}.
\ee 
   
This demonstration of self-consistency was given already in \cite{K72}, but it is sometimes discussed only in the inertial frame (e.g. \cite{BT87},\cite{BT08}). We have presented the argument here because the self-similar approach in a rotating frame gives a simpler, more inevitable description. For example, the radial and azimuthal dispersions are obviously equal in this frame and the mean rotational velocity is immediate. We also emphasize the connection with the distribution function of the scale-free  disc, and in fact with the distribution function for radial orbits in a logarithmic potential \cite{LeDHM11}, which is also of the form (\ref{eq:intF}) in 3D (the Fridmann-Polyachenko \cite{F-P84} type).

We emphasize once again that to obtain the self-similar disc from the DF (\ref{eq:intF}) one may take $E_o=k^2(\Omega_c^2-\Omega^2)r^2/2$ in equation (\ref{eq:sigmaK}). There then follows the scale-free surface density as
\be
\sigma_{SS}=4\pi K\sqrt{k^2-1}\sqrt{\frac{\Omega_c^2-\Omega^2}{2}}r.\label{eq:sigmaKSS}
\ee 
But as we shall see below, the self-similar disc must also be truncated at finite radius, which causes the potential to differ from the quadratic form. Even worse, such an isolated disc can not be in gravitational equilibrium. This must be corrected back to the self-similar form by an appropriate halo mass distribution for the self-similarity to be physically realizeable. The necessary halo is itself of some physical interest however, since rigidly rotating structures do exist in the bulges of spiral galaxies.

It is interesting to demonstrate as we remarked previously that assuming $E_I/(\Omega j_z)$ is constant suffices to pick the Fridmann-Polyachenko (or Kalnajs) DF from the family of axi-symmetric possibilities (\ref{eq:intaxiF}). For when the DF is a function only of $E>0$ we have the functional form (Kalnajs, 1976) for the DF as 
\be
F(\Phi_J)=-\frac{1}{2\pi}\frac{d\sigma}{d\Phi_J}.
\ee
Recalling equations (\ref{eq:effpot}),(\ref{eq:KE_o}),(\ref{eq:KK}) and (\ref{eq:Maclaurin}) this expression yields
\be
F(\Phi_J)=\frac{K}{(E_o-\Phi_J)^{1/2}},
\ee
and hence agrees with the functional form $F(E)$ of (\ref{eq:intF}). The argument can not be used in this form for the self-similar limit of this DF, because of the dependence of the upper limit on $r$ and hence on $\Phi_J$.

\subsection{Self-Similar Spiral Decomposition}

We start with the Poisson integral over the surface density for the disc potential, and apply a self-similar analysis. We do not now assume uniform rotation so that scaling in time $\alpha$ is generally not equal to zero. The usual dimensional analysis and Lie group formalism yields (note that $-2\delta+\mu\leftarrow \delta-2\alpha$ because of the invariance of $G$ under the scaling) 
\bea
\Phi=\Psi(\xi)e^{2(\delta-\alpha)R},& & \sigma=\Sigma(\xi)e^{(\delta-2\alpha)R}\nonumber \\
\delta r = e^{\delta R}, & & \vec{v}=e^{(\delta-\alpha)R}\vec{Y}.\label{eq:genscale}
\eea

We substitute these into the Poisson integral to obtain a relation between the scaled (also known as `reduced') potential and surface density as 
\be
\Psi(\xi)=-\frac{G}{\delta}\int_{-\infty}^\infty~du~e^{\beta u}~\int_0^{2\pi}~\frac{\Sigma(\xi')}{\sqrt{(2\cosh(u)-2\cos(\phi'-\phi))}}d\phi',\label{eq:Poisson1}
\ee
where
\bea
u\equiv \delta(R'-R),& & \xi'\equiv \phi'+\epsilon R'\nonumber\\
\beta\equiv \frac{5}{2}-2a,& & a\equiv \frac{\alpha}{\delta}.\label{eq;defs}
\eea

The case studied by Kalnajs (ibid) and presented in \cite{BT08} is in fact the only strictly self-similar case (see also \cite{SS96}), since then $\beta=0$ and the integral converges with the infinite limits. Otherwise finite limits are necessary which introduce a dependence on $R$. 

We have nevertheless introduced this slight generalization because to the extent that $R$ is small compared to say the radius of a finite disc $A$, the upper limit is then approximately independent of $R$. Then for convergence $\beta> -1/2$ ($a<3/2$) is required. This allows another class of approximately self-similar discs to be found, but without the $z$ dependence of the potential we are unable to impose self-consistency with the DF (see below). An infinite disc with a central `cut-out' would require $\beta\le 1/2$ ($a>1$) while a finite disc with a central cut-out allows general $\beta$. The integral is difficult in such cases, however.

The Kalnajs case requires $\beta=0$ and so we infer the corresponding self-similar `class' \cite{CH91} as
\be
a\equiv \frac{\alpha}{\delta}=\frac{5}{4}.\label{eq:SSclass}
\ee
Using these values in equation (\ref{eq:genscale}) reveals that $\Psi$ and $\Sigma$ are identical to the so-called `reduced quantities' used in \cite{K71} and \cite{BT08}. Kalnajs \cite{K71}, referring to earlier work by Snow, may have been aware of the Lie self-similar nature of the transformation to logarithmic radius and reduced variables. However the idea was used neither generally nor explicitly as we have done here. The scaling was discussed in a similar but more ad hoc fashion in \cite{SS96}, and their parameter $\beta=1/4$ is equivalent to the self-similar limit above. Our $a=1$ class would correspond to their class $0$, which we see would give a divergent potential unless the disc is finite. We do not discuss this case here as it is not uniformly rotating, but later in this series we study it as  the general type of `Mestel disc'.

To proceed we seek, following \cite{K71}, to express equation (\ref{eq:Poisson1}) in terms of spiral modes. We write this following \cite{BT08}\footnote{ There is an unimportant difference since to maintain formal self-similarity $\xi$ must be the independent variable so that $m$ also multiplies the term in radius}) as
\be
\Psi_m(\xi)=-\frac{G\Sigma_m}{\delta}e^{im\xi}\int_{-\infty}^\infty~du\int_0^{2\pi}~d\zeta~\frac{e^{im(\zeta+\frac{\epsilon}{\delta}u)}}{\sqrt{2cosh(u)-2cos(\zeta)}},\label{eq:Poisson2}
\ee
where $\Sigma(\xi')=\Sigma_m e^{im\xi'}$ and $\zeta\equiv \phi'-\phi$. The double integral, say $N(m,\epsilon/\delta)$, may be evaluated according to \cite{K71} as 
\be 
N(m,\epsilon/\delta)=\pi\frac{|\Gamma(\frac{1}{4}+\frac{m}{2}(1+i\frac{\epsilon}{\delta}))|^2}{|\Gamma(\frac{3}{4}+\frac{m}{2}(1+i\frac{\epsilon}{\delta}))|^2}.\label{eq:N}
\ee
We have verified this result numerically to less than one half percent in particular case with $m=2$ and $\epsilon/\delta=1$.

Although this result is an elegant derivation of a potential/density spiral pair in the plane, it does not help us to produce self-consistent spirals since we do not have the $z$ dependence. 
This can be resolved by finding the solution of the Laplace equation above the disc. We leave this treatment mostly to a subsequent work however, since such a disc does not fall into the category of uniformly rotating discs.  

We have carried out  an analysis similar to that given above for the uniformly rotating self-similar disc, which will be reported in detail elsewhere. It is perhaps worth recording here nevertheless that we find the  DF that corresponds to the disc of self-similar class given in equation (\ref{eq:SSclass}) to be  
\be
F=K(E)^2.\label{eq:KclassDF}
\ee
This gives another example where the upper limit in an integration over energy must be compatible with the self-similar class. This implies that an upper limit $E_o$ must have the form $kr^{-1/2}$. One readily finds that the integral of the DF (\ref{eq:KclassDF}) over velocity space with such an upper limit does give $\sigma\propto r^{-3/2}$ as required.

In the next section we treat the axially symmetric scale-free rigidly rotating disc. It is an approximation to a finite disc given an appropriate stabilizing halo in a sense that we study in some detail. We may, when the disc is not wholly composed of spiral arms, add this azimuthally uniform background to the non-axi-symmetric case. The non-axially symmetric potential thus acts as a non-linear perturbation on the axially symmetric background. 

\section{Axially Symmetric, Scale Free, Rigidly Rotating Disc} 

In this section we discuss the problem of stabilizing a self-similar disc by a passive halo of stars (bulge) or dark matter. For this to be possible the disc must be finite. This requirement is not removed even by taking the `disc' to be composed solely of a small number of spiral arms.  

We start by calculating the potential at radius $r$ in an axi-symmetric disc of finite radius $R$ that has  the expected surface mass density $\sigma=\Sigma~\delta r$ ($\Sigma$ constant). We do this by using the result found in equation (2.155) of \cite{BT08} to write
\be
\Phi_d=-4G\Sigma\delta\int_a^r~\frac{da}{\sqrt{r^2-a^2}}~\int_a^R~dr'~\frac{r'^2}{\sqrt{r'^2-a^2}}.\label{eq:finpot1}
\ee   
The inner integral is straight forward and after some rearrangement we obtain
\be
\Phi_d=2G\Sigma\delta R^2\int_0^{r/R}~\frac{dx}{\sqrt{(r/R)^2-x^2}}\left(x^2\ln{\left(\frac{x}{1+\sqrt{1-x^2}}\right)}-\sqrt{1-x^2}\right),\label{eq:finpot2}
\ee
where $x\equiv a/R$.

\begin{figure}
\begin{tabular}{cc} 
\rotatebox{0}{\scalebox{.4} 
{\includegraphics{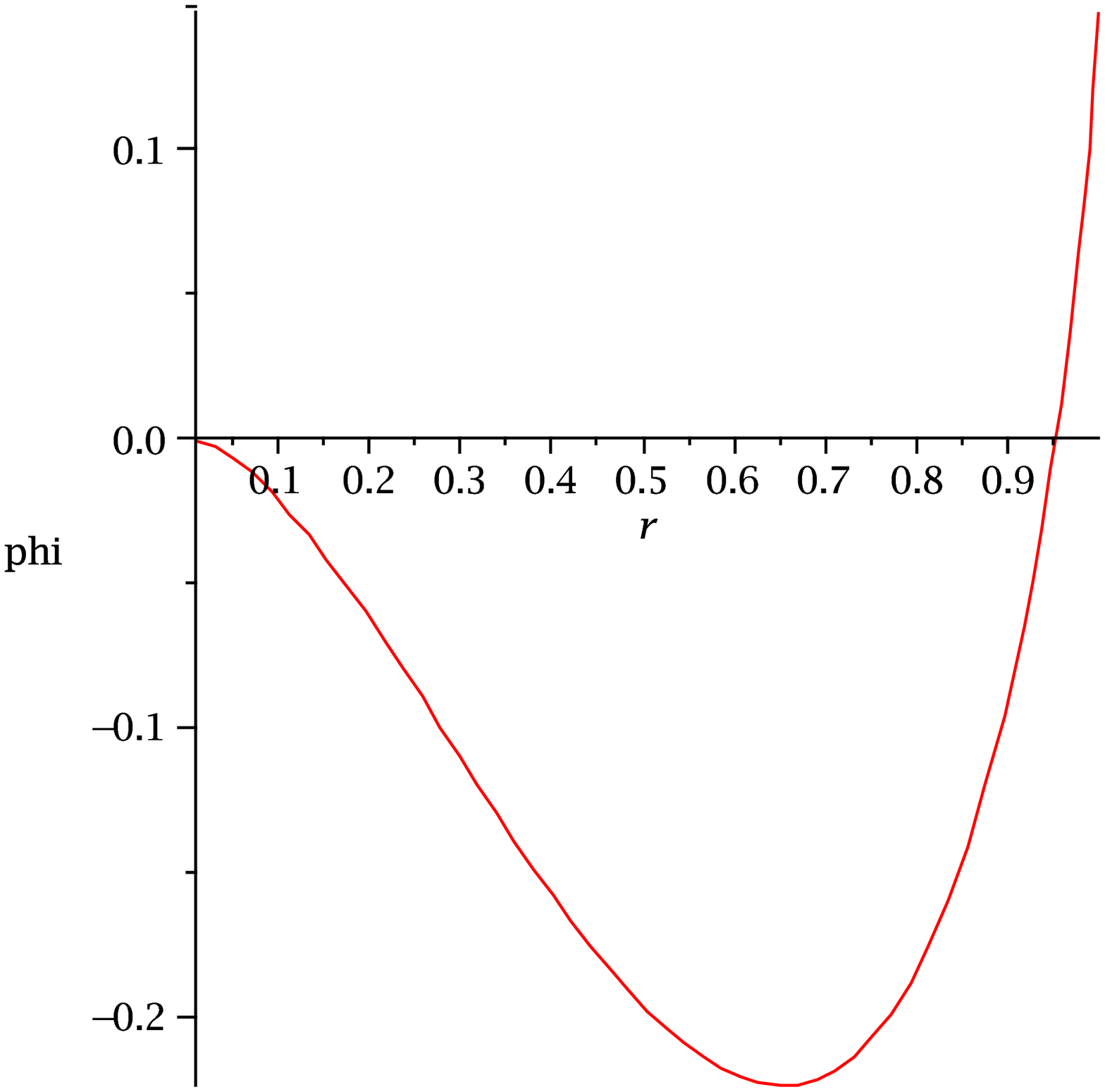}}}&
\rotatebox{0}{\scalebox{.4} 
{\includegraphics{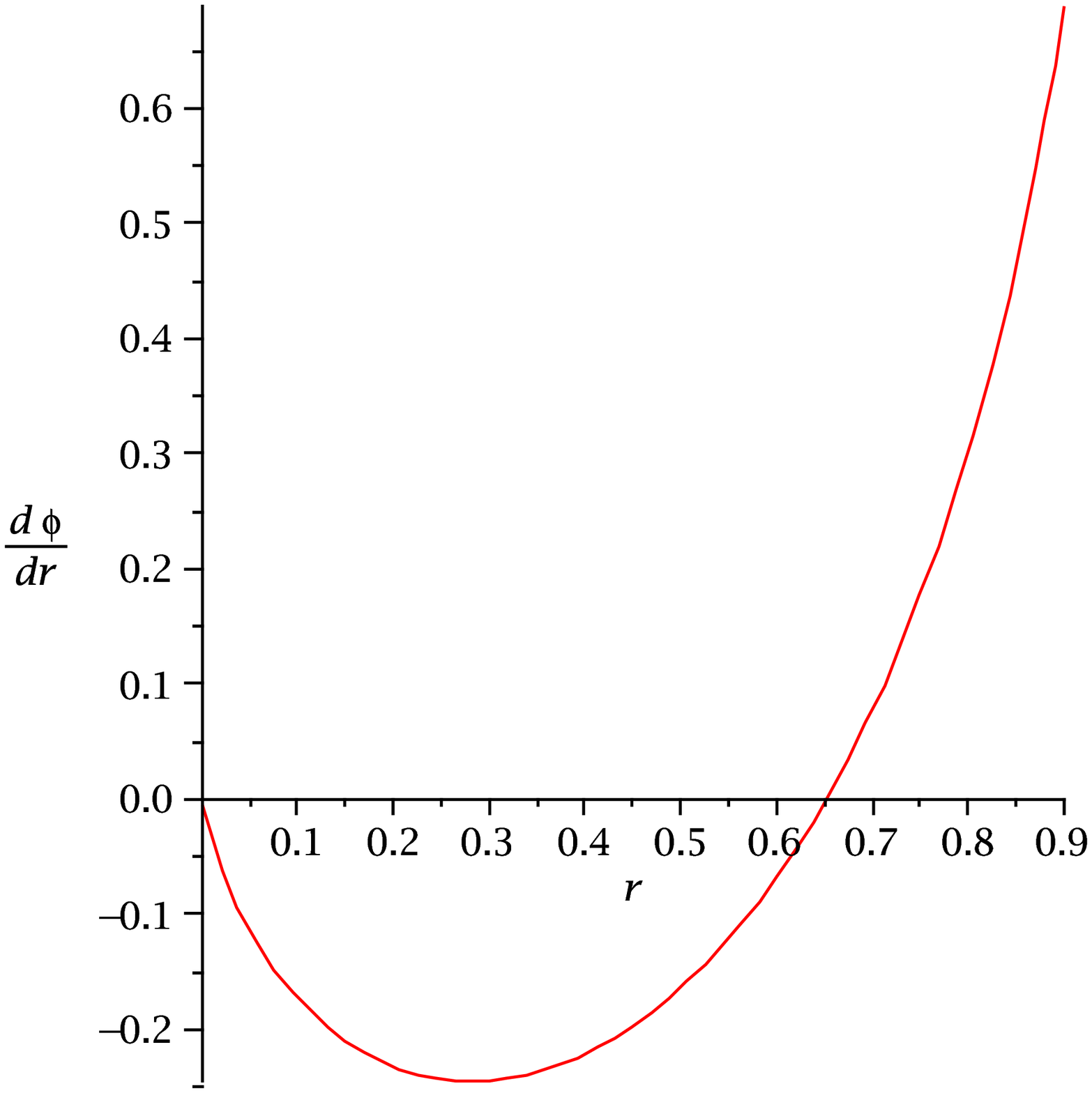}}}\\
\rotatebox{0}{\scalebox{.4} 
{\includegraphics{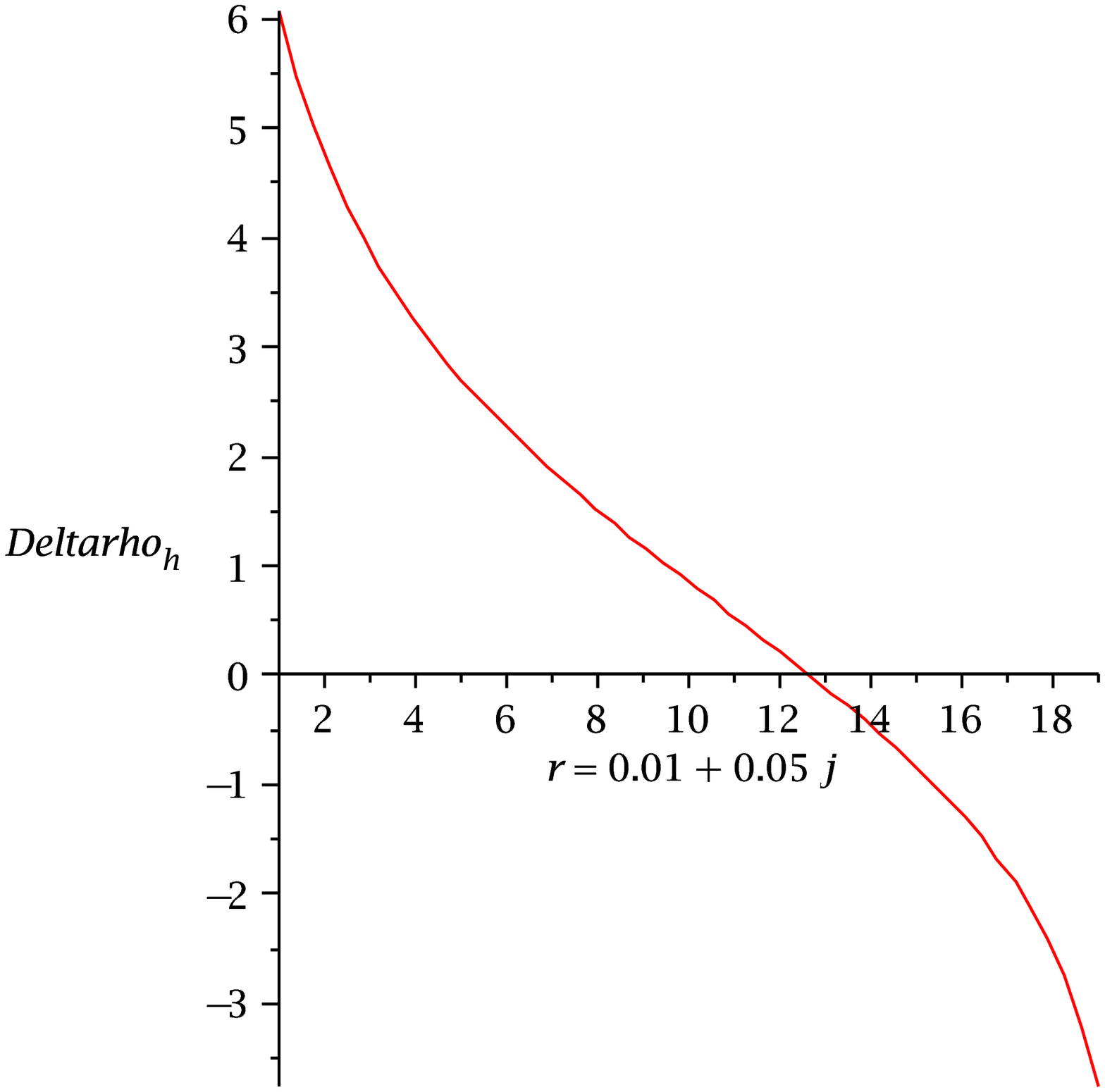}}}
\end{tabular}
\caption{ The figure at upper left shows the truncated self-similar disc potential $\Phi_d(r)$ in units of $2G\Sigma\delta R^2$ as a function of $r\equiv r/R$. We see that its gradient changes from negative to positive at $r=0.65$ so that the disc is unstable to ring formation. This is seen more directly on the plot of the potential gradient $d\Phi_d/dr$ at upper right. At lower left is shown the required perturbation to a uniform cored halo, assuming that the halo remains spherical. The zero of the potential has been placed at the origin as is the case for the self-similar potential.  }    
\label{fig:finite}
\end{figure}

To obtain the gravitational acceleration we proceed similarly by using equation (2.157) as given in \cite{BT08}. This allows us to write the gravitational acceleration in the disc at radius $r$ as
\be
-\frac{d\Phi_d}{dr}=-\frac{4G\delta\Sigma R^2}{r}\int_0^{r/R}~dx\frac{x}{\sqrt{((r/R)^2-x^2)}}\left(x\ln(\frac{x}{1+\sqrt{(1-x^2)}})+\frac{x}{\sqrt{(1-x^2)}}\right).\label{eq:finforce1}
\ee
The integrals in these two expressions are plotted in figure (\ref{fig:finite}) as a function of $r/R$.

We observe from these figures that the gravitational force is outwards for $r/R<0.65$ and inwards nearer the edge of the disc. Thus the disc is unstable to forming a very thin ring at the transition radius! Moreover the potential shows no tendency to behave self-similarly even though we have forced it to be zero at the origin. There is a small region between $0.65$ and about $0.85$ where the potential can be fitted with a quadratic law, but this is nevertheless in the context of a globally unstable disc.

In order to stabilize this behaviour while maintaining self-similarity we must require a halo to provide a potential $\Phi_h(r)$ at the disc such that 
\be
\Phi_h(r)+\Phi_d(r)=\Psi_a (\delta r)^2.\label{eq:halodiscpot}
\ee
The disc will in principle ensure that the total halo potential is not spherical \footnote{It is difficult to know this disc potential for $z\ne 0$ but we find a near-disc approximation below.} but assuming that the halo dominates, it should remain roughly spherical. Applying the Laplacian to equation (\ref{eq:halodiscpot}) yields roughly the halo density $\rho_h$ as 

\be
4\pi G \rho_h=6\delta^2\Psi_a-\frac{2}{r}\frac{d\Phi_d}{dr}-\frac{d^2\Phi_d}{dr^2}.\label{eq:halodens}
\ee

The last two terms in this expression give the spherical approximation to the non-uniform core formed from the undisturbed halo. We may refer to them as $`Deltarho_h'$ and their form in units of $r=r/R$ is shown on the lower left of figure (\ref{fig:finite}). Note that for technical reasons $r$ is written in the discrete form $0.01+0.05j$ where $j$ runs from $1$ to $19$. We see that the presence of the disc establishes a very weak cusp in the otherwise uniform (cored) halo.

This evaluation of the disturbed halo density is a crude approximation, provided that the disc contribution to the total halo potential taken spherical remains small. This will not be the case as the centre of the system is approached arbitrarily closely, because of the weak cusp. However the cusp is weaker than logarithmic (which is compatible with the variation found in numerical simulations and some theoretical estimates \cite{H07}). Given that uniformly rotating barred and spiral structures do exist in the centres of galaxies, it may be worth taking semi-seriously this form of the cusped halo. When the non-axi-symmetric structure is inside a ring, even the outwardly increasing surface density may be relevant.

In order to estimate the potential disturbance produced in the halo by a truncated self-similar disc near its centre and not too far above the disc, we take the self-similar forms literally. Then  
with axial symmetry and  $E/\Omega j_z$ constant, the  DF is the self-similar result of equation  (\ref{eq:intssP}). The Laplace equation (\ref{eq:sslaplace}) becomes simply 
\be
\partial_Z^2\Psi+4\Psi=0,
\ee
whence
\be
\Psi=A\cos{2Z}+B\sin{2Z},\label{eq:axiPsi}
\ee
where $A$ and $B$ are constants. In the disc we have that the effective potential $\Psi_J(0)=A-\Omega^2/(2\delta^2)$. It seems that this should be $\ge 0$ since otherwise gravity could not balance the mean circular velocity in the inertial frame. In this case the boundary condition (\ref{eq:ssboundary2}) yields for the axi-symmetric surface density 
\be
\Sigma_a=4\pi K(\sqrt{{\cal E}_o}-\sqrt{\Psi_J(0)}),\label{eq:axibound2}
\ee
that is explicitly
\be
\Sigma_a=4\pi K(\sqrt{{\cal E}_o}-\sqrt{(A-\frac{\Omega^2}{2\delta^2})}.
\label{eq:exaxibound2}
\ee
The energy ${\cal E}_o$  is simply an upper bound to the self-similar particle energy. Its physical correspondant scales with $r^2$.

The remaining boundary condition (\ref{eq:ssboundary1}) takes the explicit form
\be
 B=\frac{\pi G}{\delta}\Sigma_a.
\ee

For fixed $\Sigma_a$ and ${\cal E}_o$ the constants $A$ and $K$ appear to be degenerate. However if, as for the Kalnajs disc, we require gravity to give the required centripetal acceleration for the mean circular velocity, then $A=\Omega^2/(2\delta^2)$ and $\Psi_J(o)=0$. The potential is then determined everywhere as 
\be
\Psi=\frac{\Omega^2}{2\delta^2}\cos{2Z}+\frac{\pi G}{\delta}\Sigma_a\sin{2Z},\label{eq:ssaxipot}
\ee
or in terms of non-scale invariant quantities
\be
\Phi=\frac{\Omega^2 r^2}{2}\cos{\frac{2z}{r}}+\pi G\Sigma_a \delta r^2\sin{\frac{2z}{r}}.\label{eq:axipot}
\ee

When the disc is stabilized by a halo potential $A$ is equal to $(\delta r)^2\Psi_a$ where $\Psi_a$ is due to the  halo plus disc. 
For completeness we note that the self-consistent  DF amplitude is given in general by equation (\ref{eq:exaxibound2}).

The potential (\ref{eq:axipot}) is pathological as $r\rightarrow 0$ or as $z\rightarrow \infty$ but it is well-behaved on radii with $z/r$ constant, except for the divergence at radial infinity. Similar problems arise in spherical symmetry, including radial divergence and a logarithmic singularity at any point on the polar axis.
It is likely that, when using this infinite solution to approximate a finite disc of radius $R$, we will require $z\ll R$ and so the pathology will not appear off  the axis. 

Equation (\ref{eq:axipot}) does not allow consistently for the finiteness of the disc. It merely assumes finite behaviour in the self-similar form. Nevertheless its form suggests that there is only a weak dependence on the spherical angle $\pi/2-\theta=arctan(z/r)$ near the disc.

A  uniform density halo obeys the $\alpha=0$ self-similarity \footnote{In 3D the density dimension co-vector${\bf d}_\rho=(0,-3,1)\rightarrow (0,0)$ so that the self-similar density is uniform.} including the quadratic potential, so  a uniformly rotating disc and a uniform halo are of the same self-similar class. The halo is not expected to be uniformly rotating so that it must acquire its self-similarity from the boundary condition at the disc. However we have seen that the stable existence of the disc requires a distortion of the halo that is not self-similar. Provided that this distortion is small, the common self-similarity class plus the need for disc stabilization provides a kind of `disc/halo conspiracy' on the scale of the galactic nucleus.

From this section we conclude that the self-similar forms do have some application to the central parts of finite discs. In the next section we proceed therefore to give a description of the spiral structure of a uniformly rotating self-similar disc.

\section{Spiral/Bar Structure in the Rigidly Rotating Disc}

All that is new in this section stems from the solution for the non axi-symmetric potential that is found from equation (\ref{eq:sslaplace}). This solution we can expect to be superimposed on the axisymmetric solution of the last section. In effect we are treating a non-linear example of the Lin-Shu hypothesis (see e.g. \cite{BT08}) where the pattern speed is the uniform value $\Omega$ and the spiral form is stationary in the rotating frame.

Equation (\ref{eq:sslaplace}) is easily separable and readily yields a solution in the real domain. Unfortunately such a solution is not strictly periodic in the plane. It is  a Laplace transform mode in $z$ rather than a Fourier mode in the plane. This leads to discontinuities in the surface density and in the potential. 
Nevertheless, because the analysis is  transparent relative to the periodic discussion that follows, and because the discontinuities may well imitate line shocks in the plane, we discuss this case in the next sub-section. Such shock structure is probably required to select preferentially either a trailing or a leading spiral, and so to escape the `anti-spiral' theorem. 

These shocks would be in collisionless matter and although they are usually associated with plasmas they have been suggested for the rotating gravitational N-body problem since the collective interactions are similar\cite{F-P-II84},\cite{FPKL96},\cite{BT08},\cite{MV09}. 

The last reference in this series argues that non-linear Landau damping leads to small scale filamentation in phase space, which might signal the transition to a new equilibrium state. The first and second references in this list deal specifically with rotating systems and suggest the transition from soliton to shock by non-linear dissipation. The thickness of the shock is in any case finite (less than an epicyclic radius in the soliton mechanism and perhaps slightly greater than some effective Jeans length). 

Moreover there is an associated gravitational potential gradient that leads to a change in the potential across the dissipative region. Seen macroscopically after coarse graining, this implies a jump in the potential across the shock discontinuity. This process may be a reasonable description of the collisionless transition to a spiral arm, and it could either initiate or be initiated by a hydrodynamic shock. We employ this shock hypothesis in the next section, but subsequently we return to  periodic arms without shocks or distinction between leading and trailing examples.     

\subsection{Aperiodic Spiral Structure}

We separate the equation (\ref{eq:sslaplace}) according to $\Psi=S(\xi)\zeta(Z)$,and we choose $\zeta\propto \exp{(-k|Z|)}$ (where $k>0$). Moreover to remain as close to a pure planar mode as the equation permits we can  set for integer $m$
\be
k^2\equiv k_m^2=m^2(1+\frac{\epsilon^2}{\delta^2})-\frac{4}{1+\frac{\epsilon^2}{\delta^2}},\label{eq:k(m)}
\ee
as this choice makes $S(\xi)$ periodic except for the damping (equation(\ref{eq:asympot}) below).
This imposition requires $\epsilon^2/\delta^2>1$ for $m=1$. Thus the one-armed spiral `shock' (see below) is relatively tightly wound, as this requires the `winding angle' $\arctan{|\epsilon/\delta|}$ to be greater than $45^\circ$. For $m=2$ or greater the only requirement on the winding angle is that $\epsilon^2/\delta^2>0$.

The solution for each `mode' of the scaled potential becomes now
\be
\Psi_m(Z,\xi)=A_m\exp{\big[-\big (k_m|Z|+\frac{2\epsilon/\delta}{1+\epsilon^2/\delta^2}~\xi\big)\big]}\cos{(m\xi+\Theta_m)},\label{eq:asympot}
\ee
where $A_m$ and $\Theta_m$ are real constants.
 This solution has the form of a damped harmonic oscillator and so is path dependent with regard to the variation in $\xi$. This is a kind of irreversibility disguised as non-integrability. 

To complete the solution we must satisfy the boundary conditions (\ref{eq:ssboundary1}) and (\ref{eq:ssboundary2}) at the disc. The first condition gives  the form of the surface density as 
\bea
\Sigma(0,\xi)&=&\Sigma_a+\Sigma_m\equiv \Sigma_a-k_m(\delta/2\pi G)\Psi_m(0,\xi)=\nonumber\\
\Sigma_a&-&\frac{\delta}{2\pi G}~k_mA_m\exp{\big[-\big(\frac{2\epsilon/\delta}{1+\epsilon^2/\delta^2}~\xi\big)\big]}~\cos{(m\xi+\Theta_m)}\label{eq:asymdens}, 
\eea
where the potential is evaluated at $Z=0$. The surface density $\Sigma_a
$ describes an axi-symmetric component. This surface density corresponds to the net stabilized axi-symmetric potential $(\delta r)^2\Psi_a$.  

The second boundary condition  gives an expression for the surface density of collisionless matter.  We assume that the net axi-symmetric potential is positive in the inertial frame for stability, and require  that $\Psi_m(0,\xi)+\Psi_a=\Omega^2/(2\delta^2)+\Psi_m(0,\xi)>0$. In the rotating frame we suppose the net potential energy $\Psi_a+\Psi_m(0,\xi)-\Omega^2/(2\delta^2)\equiv \Psi_m(0,\xi)$ of the spiral disturbance to be negative in the peak density regions. 

We might very well assume that the peak energy is positive. With the negative assumption the perturbed gravitational force is directed outwards in the rotating frame and so particles might be expected to cross the arm on the outward leg of their `epi-cyclic' motion. With the positive assumption they would cross the arm on the inward leg. This suggests that the arm has a forward facing shock while the negative assumption implies a backward facing shock. We proceed with the negative assumption, but it should become clear how to deal with the positive energy assumption.  

We  obtain from equations (\ref{eq:ssboundary2}) and (\ref{eq:intssP}) that in regions where the spiral energy is negative
\be
f_a\Sigma_a+f_m\Sigma_m(0,\xi)=4\pi K|\Psi_m(0,\xi)|^{1/2},\label{eq:negboundary}
\ee
where $f_i<1$ is the fraction of the disc material $\{i\}$ that is collisionless. The origin of $(1-f_i)\Sigma_i$ must be in another form of matter, almost certainly gas and dust in a spiral galaxy. It is of course possible that $f_a=f_m$.

We substitute for the spiral surface density on the left of this last equation using the appropriate part of equation (\ref{eq:asymdens}). In this way we obtain a condition for self-consistency for each value of $m$ in the negative energy region. Strictly one should only apply  this condition when it contains the disturbed potential summed over all $m$, but in view of the nearly pure modal behaviour of at least the `grand design' spiral galaxies we explore each mode independently.

In regions of the disc where the spiral energy is positive the right hand side of equation (\ref{eq:negboundary}) must be replaced by $4\pi K(\sqrt{{\cal E}_o}-\sqrt{\Psi_m(0,\xi)})$ as per our previous discussion. An essential difference in this non-axially symmetric disturbance is that ${\cal E}_o$ can be a function of $\xi$, without destroying the self-similarity. In fact in order to pass smoothly through the zeros of $\Psi_m(0,\xi)$ it should have the form $k^2\cos{(m\xi+\Theta_m)}$ in the positive potential regions.  

The self-consistency condition in the negative energy regions becomes explicitly
\bea
&f_m&\left[-\frac{\delta}{2\pi G}k_mA_m\exp{\big [-\big (\frac{\epsilon/\delta}{1+\epsilon^2/\delta^2}\big )~\xi~\big ]}~\cos{(m\xi+\Theta_m)}\right] +f_a\Sigma_a \nonumber\\  
&=&4\pi K\big |A_m\exp{\big [-\big (\frac{2\epsilon/\delta}{1+\epsilon^2/\delta^2}\big )~\xi~\big ]}~\cos{(m\xi+\Theta_m)}\big |^{1/2}.\label{eq:aperselfcon}
\eea
Dimensionally we see that everything is coherent so long as $A_m$ has the dimensions of squared velocity. 

There are various ways to satisfy this equation that we now consider.
It is apparent that one choice of modal constants $\{A_m,\Theta_m\}$ will not allow this condition to be satisfied for all $\xi$, for constant $K$, $\Sigma_a$ and $f_i$. It is possible to see this condition as an equation for the single fraction $f(\xi)=f_i$ provided that its value remains less than one. However this leaves much arbitrariness in the problem. In order to obtain discrete spiral arms we shall require that $f_i\le 1$ be  known values on the arm\footnote{Any fixed values would serve, but ultimately they are likely to be fixed by observation.}. This will determine the arm.  Since then the condition (\ref{eq:aperselfcon}) can only be satisfied at discrete values of $\xi$ by fixing the modal constants. The two dimensional nature of the analysis therefore implies that particles in the  background pass to and from the discrete spiral arms. 
 
One does not need to satisfy the second boundary condition outside of the spiral arms but it is likely to apply to the collisionless part of the disc. One must then  treat the equation (\ref{eq:aperselfcon}) as an equation for $f(\xi)$ once we add the information that $f_a=f_m=f$. This means that $f$ will be discontinuous across the arms.

To illustrate the possible solutions, we consider first the one-armed mode with $m=1$. Let us place a discrete spiral arm at $\xi=0$ by choosing $A_1>0$ and $\Theta_1=\pi$, so that we have a positive spiral density on the background according to equation (\ref{eq:asymdens}). We recall that in this case we are restricted to $\epsilon/\delta>1$, so that the one-armed winding angle is always greater than $45^\circ$. Then equation (\ref{eq:aperselfcon}) determines $A_1$ from

\be
 A_1^2+(2\eta^2-\nu^2/f^2)A_1-\eta^4=0,\label{eq:A1}
\ee

where 
\be
\nu\equiv \frac{8\pi^2GK}{k_1\delta},~~~~~\eta^2\equiv \frac{2\pi G\Sigma_a}{\delta k_1}, \label{eq:nu}
\ee
and $\nu,\eta$ have the dimensions of velocity. This equation always has one positive root when $\nu^2>4\eta^2f^2$. All of the modal constants are now fixed by choosing $f\le 1$. 

Starting at the positive or leading `side' of the $\xi=0$ spiral and incrementing $\phi$ by $2\pi$ at a fixed $R$ brings us to the negative or trailing side with a density lower by a factor $\exp{(-2\pi b)}$. We have defined for convenience 
\be
b\equiv \frac{2\epsilon/\delta}{1+\epsilon^2/\delta^2},\label{eq:b}
\ee
which is always less than or equal to $1$. 
There is thus a backward facing `shock' type discontinuity from the trailing to the leading edge of the spiral. The spiral potential has the more negative value by the factor $\exp{(2\pi b)}$ after a $2\pi$ increase in $\phi$ at fixed $R$. This is really a coarse-grained description of the potential gradient in a collisionless layer or shock.

\begin{figure}
\begin{tabular}{cc} 
\rotatebox{0}{\scalebox{.4} 
{\includegraphics{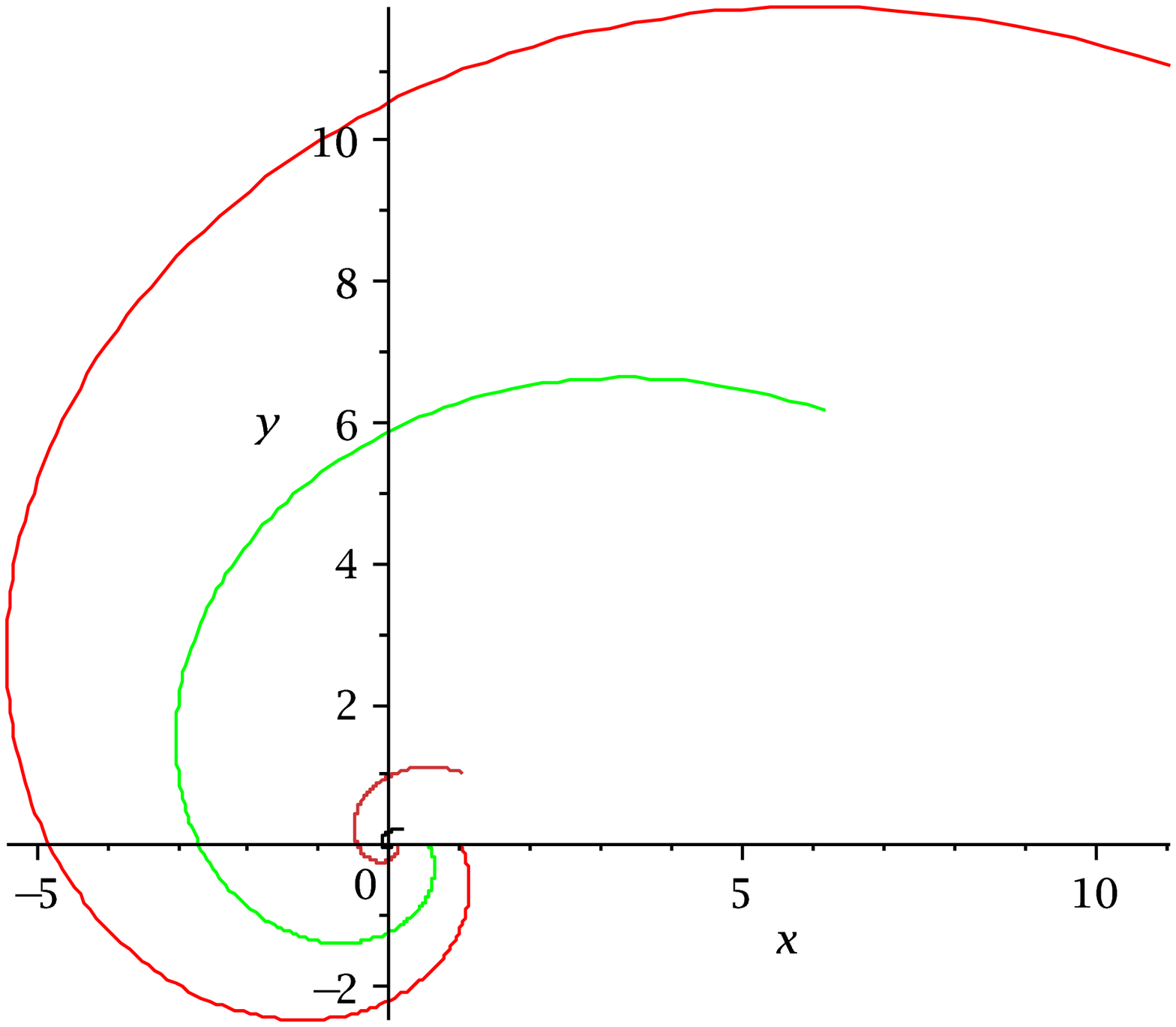}}}&
\rotatebox{0}{\scalebox{.5} 
{\includegraphics{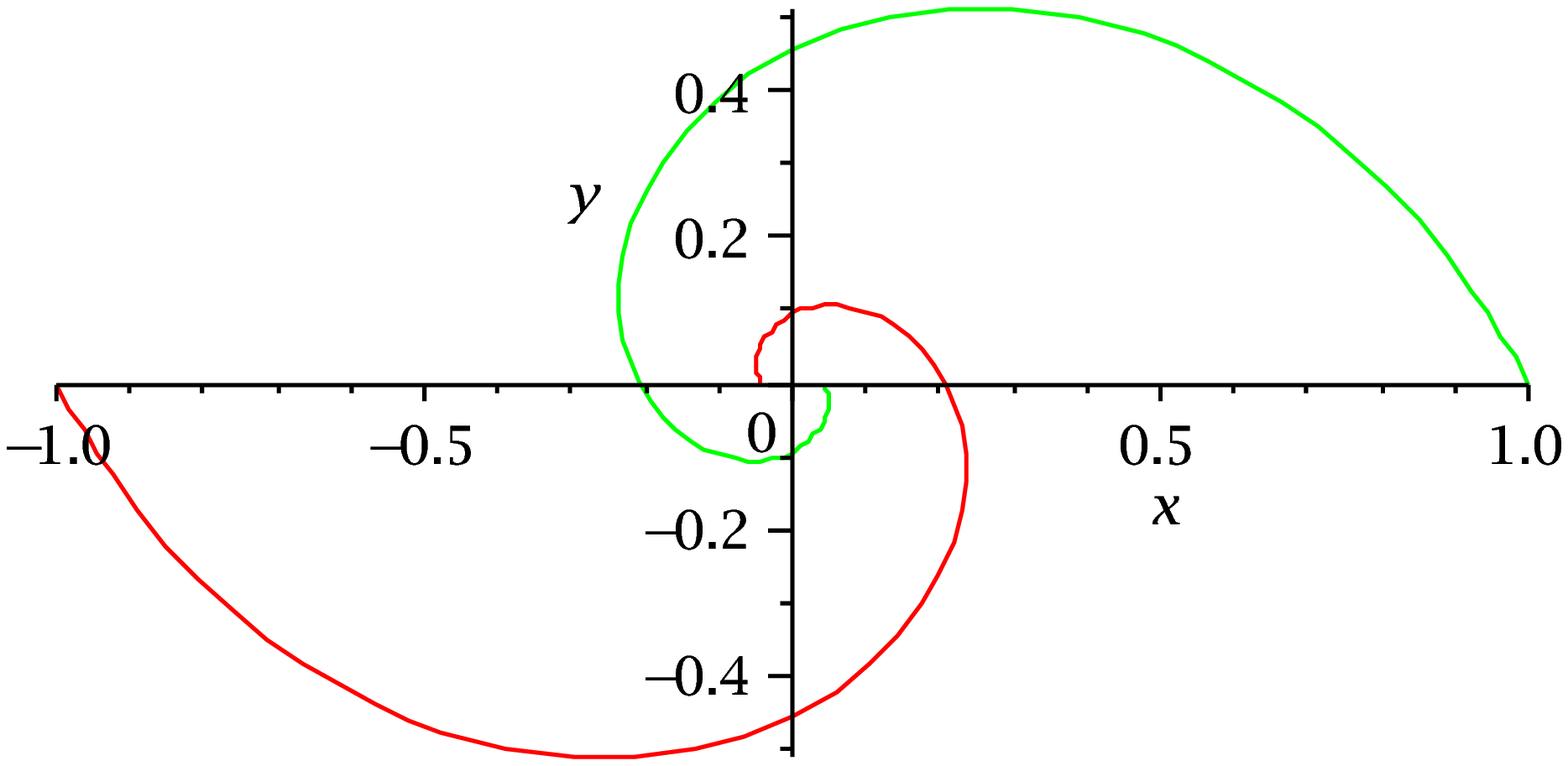}}}\\
\rotatebox{0}{\scalebox{.4} 
{\includegraphics{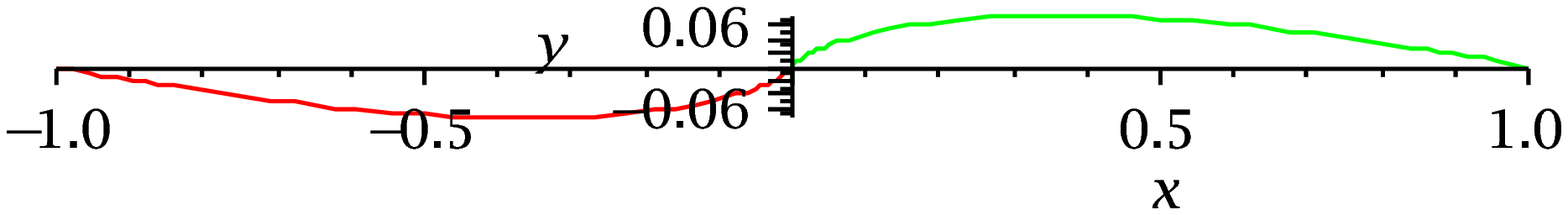}}}
\end{tabular}
\caption{The figure shows on the left a polar plot of one-armed aperiodic spirals that are solutions to equation (\ref{eq:roots1}) with $\epsilon/\delta=2$. In descending order of size the spirals are at $\xi=\{0,-1.1666,-4.7350,-7.8521\}$ respectively. One sees that they are separated by about $\pi$ along the $y=0$ axis. The two-armed spiral on the right has discrete arms at $\xi=0$ (right spiral at $y=0$) and at $\xi=\pi$ (left spiral at $y=0$). The lower figure shows the same two-armed spiral when $\epsilon/\delta=0.2$, giving a bar-like structure. In all cases the stellar motions should be along the arms for physical consistency.  }    
\label{fig:m1-2spirals}
\end{figure}

There is another type of solution that gives a set of discrete spirals.
The same value for $A_1$ given by equation (\ref{eq:A1}) will apply on discrete spirals $\xi_s$ that satisfy 
\be
e^{-b\xi_s}\cos{\xi_s}=1.\label{eq:roots1}
\ee
This equation has an infinite number of negative roots which, after a least negative root, oscillate in $\xi_s$  between close to $-270^\circ$ and $-90^\circ$. 
We have converted the radian measure of $\xi_s$ to degrees (modulo $360$) in order to illustrate the separation in angle at a fixed radius. For example if we choose $\epsilon/\delta=2$ so that $b=0.8$, than the first root (after zero) is $-66^\circ.843$,and the succeeding two roots are $-271^\circ.297$  and $-89^\circ.893$ (modulo $360$) respectively. These spirals are shown in figure (\ref{fig:m1-2spirals}). The fourth root spiral is too small to be seen on this scale that is chosen so that $r=1/\delta\equiv r_o$ at $\phi=0$ on the $\xi=0$ spiral.    

In this way we have constructed a set of one-armed spirals, each of which will have a density jump from the leading to the trailing side of magnitude $e^{(2\pi b)}$ for trailing spirals. The reverse is true for leading spirals.

One can also use the continuity of the collisionless fractions to determine them in terms of $A_m$ by writing equation (\ref{eq:aperselfcon}) at $\xi=0$ and at $\xi=2\pi$ and solving for the $f_i$. This yields formally
\bea
f_m&=&\frac{\nu}{\sqrt{A_m}}\frac{(1-e^{-\pi b})}{(1-e^{-2\pi b})},\nonumber\\
f_a&=& \frac{\nu}{\eta^2}~e^{-\pi b}\frac{(1-e^{-\pi b})}{(1-e^{-2\pi b})}. 
\eea
These must be less than one but $A_m$ remains arbitrary otherwise.

For the two-armed spiral with $m=2$, the same array of solutions is possible as were found for the one-armed spiral above.
We can once again start with an arm at $\xi=0$ and satisfy the equation (\ref{eq:aperselfcon}) by insisting that $A_2$ satisfy the equation (\ref{eq:A1}) wherein $k_1$ replaced by $k_2$. For symmetry with a second arm at $\xi=\pi$, we require a jump in the density between the leading and trailing sides of magnitude $e^{b\pi}$. Once again the reverse is true for leading arms. 

The main difference between the two-armed spiral and the one-armed spiral is that we are allowed to construct bars in this case by allowing $\epsilon/\delta\rightarrow 0$. On the right-hand side of figure(\ref{fig:m1-2spirals}) we show a two-armed open spiral of this type having $\epsilon/\delta=2$ together at lower left with a nearly bar-like structure having $\epsilon/\delta=0.2$.

Just as in the one-armed example we can construct a system of spiral arms located where 
\be
e^{-b\xi_s}\cos{2\xi_s}=1,\label{eq:m2structure}
\ee
although the constant on the right could be any constant $c$ that would then be absorbed into $A_2$. The effect is  to delete the arm at $\xi=0$.   
 
The roots of this last equation behave similarly to those found in the one-arm case. Using once again the value $b=2$, there is a first root after the  zero root at $\xi_1\approx- 139^\circ.12$ and then the roots oscillate (modulo $360^\circ$) between approximately $-45^\circ$ and $-225^\circ$. In fact they converge on these latter two values. Since each value of $\xi_s$ found in this manner has its counterpart at $\xi_s-\pi$, these convergent pairs are the natural two-armed spiral. To ensure this we must remove the extraneous roots at zero and the first root $\xi_1$. This can be done by setting $c>~8$ as the constant on theright of equation (\ref{eq:m2structure}). 
In this way we arrive naturally at a two-armed spiral structure. Each arm will have a density jump of $e^{a\xi}$ from the leading to the trailing side (and the reverse for leading arms).

One might think that this `bipolarity' that arises in the one-armed case and in the two-armed case is an argument for the ubiquity of two-armed spirals. However, similar behaviour occurs for $m=3$ where the interval between the convergent roots is $\approx 120^\circ$ as would be expected. The principal difference as $m$ increases is that the strength of the `shock' is  reduced (to $e^{2b\pi/3}$ for $m=3$) for the same winding angle. Thus, although the bi-polarity is not favoured by the system, the definition and hence the visibility of the arms may be maximal in the two-armed mode (after the true  one-armed system). In this connection we note again that $b(\epsilon/\delta)$ is a maximum (of one) at $\epsilon/\delta=1$ so that the strength of the discontinuity is reduced as the winding increases towards tight spirals or decreases towards bars. The interesting feature of this solution is that, because the oscillating roots become rather dense as the sequence continues, it is possible to imagine a true 2D PDF that defines the arms in this type of solution.

In fact a true 2D solution to the self-consistency condition (\ref{eq:aperselfcon}) also exists even with $f$ constant, if the amplitude of the DF is $K(\xi)$. This is degenerate with a variation in $f(\xi)$. But such a variation requires $\xi$ to be a characteristic constant of the CBE, which in turn by equation (\ref{eq:sscbe}) requires 
$\delta Y_\phi+\epsilon Y_R=0$. This is the condition that the particles in the disc in the rotating frame move purely backwards and forwards along the spiral direction, since $Y_\phi/Y_R=rd\phi/dr=-\epsilon/\delta$. Unhappily this requirement reveals $Y_R$ and $Y_\phi$ to be dependent variables, so that the Boltzmann equation (\ref{eq:sscbe}) is no longer correct. One of $Y_R$, $Y_\phi$, must be integrated out subject to the constant $\xi$ condition. This procedure is much more transparent in spiral based coordinates that we introduce below. 

The constants $f_m$ and $f_a$ could be determined also by continuity as we did for the one-armed case.
  
Having studied spirals with shock type dissipation, we turn in the next section to consider the purely periodic, modal solutions to equation (\ref{eq:sslaplace}). Although these have a familiar modal structure, they do not offer a consistent means of avoiding the anti-spiral theorem. 

\subsection{ Periodic Spiral Structure}

In this section we seek periodic solutions of equation (\ref{eq:sslaplace}) by writing 
\be
\Psi=e^{-(im \xi)}\zeta (Z),\label{eq:perisol}
\ee
from which we must extract the real part. One must sum over all positive and negative integer values of $m$ in order to describe an arbitrary periodicity, but we shall content ourselves with looking at pure modes. These modes, we recall, are in the rotating frame and are superimposed on the axi-symmetric background. By substituting the assumed form (\ref{eq:perisol}) into equation (\ref{eq:sslaplace}), we obtain an equation in the complex plane for $\zeta(Z)$ as
\be
 \frac{d^2\zeta}{dZ^2}+[4(1-im\frac{\epsilon}{\delta})-(1+\frac{\epsilon^2}{\delta^2})m^2]\zeta=0. \label{eq:modallaplace}
\ee
The potential for negative $m$ and leading spirals is thus the same as that for positive $m$ and trailing spirals.

This harmonic equation is readily solved  formally, but it can be tedious to extract the real part. It is much easier to consider individual modes. The solution in cylindrical coordinates has singularities at large $z$ and small $r$, which is an apparent argument against a modal analysis in these coordinates. However we recall that the potential is everywhere well-behaved on a cone generator of constant $Z$, which is effectively the same behaviour as is found in spherical-polars. Indeed an analysis in spherical polars does not change the essential behaviour of the stellar disc. The CBE is identical in the plane to the one we have used even when written  in rotating spherical polars. In any case we know that this analysis holds only with a stabilizing halo and  a truncated disc not too far from the plane.

Starting with the one-armed structure when $m=1$ we obtain as a solution for $\zeta(Z)$ \footnote{In the subsequent equations of this section $\epsilon\leftarrow \epsilon/\delta$ for simplicity, unless otherwise noted}
\be
  \zeta_1(Z)=C_1\sin{(\sqrt{1-4i\epsilon-\epsilon^2}~Z)}+C_2\cos{(\sqrt{1-4i\epsilon-\epsilon^2}~Z)}.
\ee
The constants $C_1$, $C_2$ are complex in general, but since in their complex form they only add a phase to the sine and cosine functions, we take them to be real for simplicity. The real part of $\Psi_1(\xi,Z)$ still possesses a formidable form, but in the plane it becomes simply 
\be
Re(\Psi_1(0,\xi))=C_2\cos{\xi}.\label{eq:peripot}
\ee
The derivative $d\Psi_1(\xi,Z)/dZ$ is slightly more involved at the plane as   
\bea
Re\left(\frac{d\Psi_1}{dZ}\right)_0&=\frac{C_1}{2}\bigr (\sqrt{2\sqrt{(\epsilon^2+9)(\epsilon^2+1)}+6-2\epsilon^2}~(\cos{\xi})\nonumber\\
&-\sqrt{2\sqrt{(\epsilon^2+9)(\epsilon^2+1)}-6+2\epsilon^2}~(\sin{\xi})~\bigr),\label{eq:peripotderiv}
\eea
which is related to the total surface density through
\be
\Sigma_1(0,\xi)=\Sigma_a+\frac{\delta}{2\pi G}Re\left(\frac{d\Psi_1}{dZ}\right)_0.\label{eq:totperidens}
\ee

We assume once again that the spiral disturbance has negative energy in the rotating frame and that $\Psi_a=\Omega^2/(2\delta^2)$ so that equation (\ref{eq:negboundary}) applies. Then the self-consistency equation becomes on using equations (\ref{eq:peripot}) and (\ref{eq:peripotderiv}) 
\bea
&f_1&\bigl[C_1\bigl(\sqrt{2\sqrt{(\epsilon^2+9)(\epsilon^2+1)}+6-2\epsilon^2}~\cos{\xi}-\sqrt{2\sqrt{(\epsilon^2+9)(\epsilon^2+1)}-6+2\epsilon^2}~\sin{\xi}\bigr)\bigr]+f_a\frac{(4\pi G\Sigma_a)}{\delta}\nonumber\\
&=&\frac{16\pi^2KG}{\delta}|C_2\cos{\xi}|^{1/2}.\label{eq:selfconperi} 
\eea

Even with a complex $C_2$ that would add a phase constant to $\xi$ under the cosine, there is no solution for the constants that would apply for all $\xi$ with $K$ and $f_i$ constant. We proceed therefore as in the previous section by setting the spiral arm at $\xi=0$ with $f_1=f_a=f$ fixed. Otherwise we suppose that equation (\ref{eq:selfconperi}) gives $f(\xi)$ in the negative energy region(and the corresponding equation in the positive energy region).  

For the arm at $\xi=0$, the constant $C_1>0$ follows by setting the desired surface density in the arm from equation (\ref{eq:totperidens}) given the winding angle. 
Subsequently, $C_2$ follows from equation (\ref{eq:selfconperi}) given $K$ and $f$. There is no restriction of winding angle and no need for shock-type discontinuities. Both trailing and leading arms are possible. Once again collisionless particles must enter and leave the arm.

For the two-armed mode with $m=2$ the solution for $\Psi_2(\xi,Z)$ is formally (the constants are now peculiar to this mode)
\be
\Psi_2=e^{-2i\xi}\left(C_1~e^{2\sqrt{\epsilon}\sqrt{(2i+\epsilon)}Z}+C_2e^{-2\sqrt{\epsilon}\sqrt{(2i+\epsilon)}Z}\right),
\ee
from which the real part at $Z=0$ follows as (we take $C_1$ and $C_2$ real)
\be
Re(\Psi_2(0))=(C_1+C_2)\cos{2\xi}.
\ee
Similarly some calculation  shows that
\be
Re(\frac{d\Psi}{dZ})_0=(C_1-C_2)\left(b_+(\epsilon)\cos{2\xi}+b_-(\epsilon)\sin{2\xi}\right).
\ee
We have set for brevity
\be
b_{\pm}=[2\epsilon\sqrt{\epsilon^2+4}\pm 2\epsilon^2]^{1/2}.
\ee
For large $\epsilon$, that is for very tightly wound spirals, $b_-\rightarrow 0$.

The self-consistency condition can now be written for negative energy spirals as
\be
f_2\bigl[(C_1-C_2)(b_+(\epsilon)\cos{2\xi}+b_-(\epsilon)\sin{2\xi})\bigr]+f_a\frac{2\pi G\Sigma_a}{\delta}=\frac{8\pi^2KG}{\delta}\sqrt{|(C_1+C_2)\cos{2\xi}|}.\label{eq:m2modselfcon}
\ee
Once again we see that the arms are discrete. They can be taken to lie at $\xi=0$ and at $\xi=\pi$, while again $f_2=f_a=f$ on the arms. Otherwise the equation determines $f(\xi)$ for the inter-arm regions as above.
 
There are two constants $C_1-C_2$ and $C_1+C_2$ to be determined. The self-consistency condition (\ref{eq:m2modselfcon}) at $\xi=0$ or $\pi$ will  determine one for fixed $K$ and $f$. However if the underlying axi-symmetric disc is given, so that  $\Sigma_a$ is known then  $C_1-C_2$ may be determined by giving the total arm density.

It is worth noting that in the rotating frame there is no net flux of particles because of the isotropic nature of the Kalnajs DF. Thus particles enter and leave the spiral arms following what would be essentially epicyclic orbits in the inertial frame in the linear theory. In the multi-arm case particles may travel between the arms. 

We turn in the next section to an analysis based on the spiral structures as coordinate lines. This enables us to easily pass to a treatment where only motion along discrete spiral arms is present.   

\section{Spiral Arm Based Analysis}

We begin this section by transforming from plane polar coordinates $\{r,\phi\}$ to spiral coordinates $\{\ell,\xi\}$, where $\ell$ is measured as length along the spiral and $\xi$ plays the r\^ole of orthogonal coordinate. This requires a careful representation of the Boltzmann equation although the Poisson equation changes very slightly.

\subsection{Equations in Spiral Coordinates}

The unit vectors locally parallel and orthogonal to a spiral arm defined by 
\be
\xi=\phi+\frac{\epsilon}{\delta}\ln{\delta r},
\ee
may be taken respectively as 
\be
\vec{e}_\ell=\frac{(\vec{e}_r-\frac{\epsilon}{\delta}\vec{e}_\phi)}{\sqrt{1+\frac{\epsilon^2}{\delta^2}}},\label{eq:epar}
\ee
and
\be
\vec{e}_\xi=\frac{(\frac{\epsilon}{\delta}\vec{e}_r+\vec{e}_\phi)}{\sqrt{1+\frac{\epsilon^2}{\delta^2}}}.\label{eq:eortho}
\ee

These lead us to introduce the corresponding velocity components
\be
v_\ell\equiv \vec{e}_\ell\cdot\vec{v},~~~~v_\xi\equiv \vec{e}_\xi\cdot\vec{v},\label{eq:velocitycomps}
\ee
and these define in turn the length along the spiral $d\ell=v_\ell(\xi)dt$ and an orthogonal velocity $r\dot\xi=r\dot\phi+(\epsilon/\delta)\dot r\equiv v_\phi+(\epsilon/\delta)v_r$. The first definition integrates to
\be
\ell=r\sqrt{1+\frac{\epsilon^2}{\delta^2}},\label{eq:ell}
\ee
while the second expression can be written using the definition of $v_\xi$ as 
\be
v_\xi=\frac{\ell\dot\xi}{(1+\frac{\epsilon^2}{\delta^2})}.
\ee
Note that we also have by definition that $v_\ell=\dot\ell$. 

We are now in a position to follow the usual route to the two dimensional CBE in a uniformly rotating frame by expressing the disc particle Lagrangian in terms of $\{\ell,\dot\ell,\dot\xi,\xi\}$ in this frame, hence passing to the Hamiltonian, and finally to $dF/dt=\partial F/\partial t+[F,H]=0$ ($[F,H]$ is the Poisson bracket). 

The resulting canonical form does not hold the velocities $\{v_\ell,v_\xi\}$ constant in the partial derivatives since the canonical momenta $\{p_\ell,p_\xi\}$ are the independent variables. Hence appropriate transformations of the partial derivatives (including to $(\partial F/\partial\ell)_{\vec{p},\xi}$) must be made to obtain the following form  
\bea
\partial_tF&+&v_\ell\partial_\ell F+\frac{(1+\frac{\epsilon^2}{\delta^2})}{\ell}v_\xi\partial_\xi F+\left(\frac{\Omega^2\ell+2\Omega v_\xi}{1+\frac{\epsilon^2}{\delta^2}}+\frac{v_\xi^2}{\ell}-\partial_\ell\Phi\right)\partial_{v_\ell}F\nonumber\\
&-& \left(\frac{2\Omega v_\ell}{1+\frac{\epsilon^2}{\delta^2}}+\frac{v_\ell v_\xi}{\ell}+\frac{1+\frac{\epsilon^2}{\delta^2}}{\ell}\partial_\xi\Phi\right)\partial_{v_\xi}F=0.\label{eq:spiralcbe}
\eea
This equation is general in a uniformly rotating frame. One sees that it reduces to equation (\ref{eq:stacbe}) as $\epsilon/\delta\rightarrow 0$ (when the time dependence is removed for stationary rotating structures) as it should. The presence of the winding angle as a parameter opens the way to an expansion of the CBE either `near' (in parameter space) bar-like structures ($\epsilon/\delta\rightarrow 0$) or `near' ring-like structures for which $\epsilon/\delta\rightarrow \infty$, but we do not pursue this here. 

The Poisson equation above the disc can  be written simply in $\{\xi,\ell,s\}$ coordinates, where $s\equiv \sqrt{1+\epsilon^2/\delta^2}z$, as 

\be
\frac{1}{\ell}\partial_\ell(\ell\partial_\ell\Phi)+\frac{1}{\ell^2}\partial_\xi^2\Phi+\partial_s^2\Phi=0.\label{eq:spiralPoiss}
\ee
This equation together with the previous equation may be used as usual for a two dimensional disc to impose self-consistency. The results do not differ of course from those found above, although the description is perhaps more natural. However whenever a single spiral arm is to be discussed, as is the case in this section, an infinite number of `modes' of this equation (to form a delta function in $\xi$) are required to satisfy correctly the boundary condition at the arm. We shall see that it is better to use the Poisson integral in such a case since then we simply bypass the difficult solution for the potential above the plane in favour of that in the disc. 

\subsection{Discrete Spiral Arms}

We wish in this section to consider the case where in the rotating frame the particles move only parallel to the spiral arm. This is true one dimensional motion in our rotating spiral coordinates. Realistically, the motion should extend over a finite region in the $\xi$ coordinate. We find that although such finite structures are not so violently unstable as is the finite two-dimensional self-similar disc, they do still require weak adjustments to the halo in order to ensure self-similarity and stability.

We assume that the collisionless distribution function on a spiral line can be written as
\be
F=F_\ell(\ell,v_\ell)\frac{\delta_D(v_\xi)\delta_D(\xi-\xi_s)}{\ell},\label{eq:1DDF}
\ee
where $\delta_D(x)$ is the Dirac delta function. An integration of equation (\ref{eq:spiralcbe}) over $v_\xi$ and $\xi$ yields 
\be
v_\ell\partial_\ell F_\ell+\left(\frac{\Omega^2\ell}{1+\frac{\epsilon^2}{\delta^2}}-\partial_\ell\Phi\right)\partial_{v_\ell}F_\ell=0,\label{eq:1DBE}
\ee
where $F_\ell$ and $\Phi$ are evaluated on the spiral $\xi=\xi_s$.

To give the spiral arm a modestly finite extent in $\xi$ we can interpret the delta function according to 
\be
\delta_D(\xi-\xi_s)=Limit_{(\xi_1,\xi_2)\rightarrow \xi_s}\left(\frac{\Theta(\xi-\xi_1)-\Theta(\xi-\xi_2)}{\xi_2-\xi_1}\right),\label{eq:deltadirac}
\ee
but replace the strict limit by a finite interval $[\xi_1,\xi_2]$. Here $\Theta(x)$ denotes the Heaviside step function, equal to $1$ when the argument is positive and zero otherwise.

We are now in a position to carry out a self-similar analysis for the DF on discrete spirals in a uniformly rotating frame of reference.  For  line masses the uniformly rotating ($\alpha=0$) self-similar mass per unit length $\lambda\propto r^2$ since ${\bf d}_\lambda=(0,-1,1)\rightarrow (0,2)$. 
 We write, following the usual technique with $\alpha=0$, 
\bea
\delta\ell=e^{\delta L}~~& &~~\delta s=Ze^{\delta L}\nonumber\\
F_\ell=P_\ell e^{\delta L}~~& &~~\Phi=e^{2\delta L}\Psi~~~~v_\ell=Y_\ell e^{\delta L}\label{eq:1Dssvars}
\eea
and substitute into equation (\ref{eq:1DBE}) to obtain 
\be
Y_L\partial_LP_\ell-(\delta Y_L^2+\partial_L\Psi+2\delta\Psi-\frac{\Omega^2}{\delta})\partial_{Y_L}P_\ell=-\delta P_\ell.\label{eq:ss1DBE}
\ee

We recall that only when $\partial_L\Psi=0$ do we have strict self-similarity, but even allowing this not to be the case  the following conclusion holds. The characteristics of the latter equation allow us to conclude that ($ds$ here is an increment along the characteristic)
\be
\frac{d{\cal E}}{ds}=-2\delta Y_L{\cal E},\label{eq:enchar}
\ee
where ${\cal E}\equiv Y_L^2/2+\Psi-\Omega^2/2\delta^2$, and 
\be
\frac{dP_\ell}{ds}=-\delta Y_L P_\ell.\label{eq:DFchar}
\ee

Together these two equations imply that the one dimensional DF on a spiral is
\be
P-\ell=K|{\cal E}|^{1/2},\label{eq:ss1DDF}
\ee
or equivalently in physical variables
\be
F=K|E|^{1/2}.\label{eq:1DphysDF}
\ee

Only when there is no self-similarity does $Y_L=dL/ds$, which allows equation (\ref{eq:enchar}) to be integrated directly to give ${\cal E}\exp{2L}=constant=E\equiv v_\ell^2/2+\Phi-\Omega^2\ell^2/2$. In that case, just as we found in our discussion of the two dimensional DF at the beginning of this paper, we might have $K=K(E)$ and so revert to the Jeans' theorem.

Equation (\ref{eq:1DphysDF}) reveals a satisfying progression in the self-similar DF with dimension for a system in uniform rotation. In one dimension it is $\propto \sqrt{|E|}$ as found here, in two dimensions it is $\propto1/\sqrt{|E|}$ as found earlier in this paper, and in three dimensions it is $\propto 1/(\sqrt{|E|})^3$. This latter result may be inferred from \cite{LeDHM11b} for the isotropic limit when $a=0$, although the energy would have to be written in the uniformly rotating frame as above. The basic result as a function of $a$ in the maximally coarse-grained steady limit may be found in earlier papers such as \cite{H07} and references therein. For strict radial infall in spherical symmetry, the self-similar DF is also found to be $\propto \sqrt{|E|}$ \cite{HW95},\cite{LeDHM11}  

We can imagine this collisionless spiral arm to be superimposed on the usual axi-symmetric, truncated self-similar disc, for which both the density and the potential (quadratic after stabilized by the halo) are known. This may be made in part of collisionless matter as above. However here we do not allow a constant interchange of particles between the arm and the disc

Just as in the two dimensional case, one  should verify to what extent a truncated self-similar discrete  arm can be stabilized by an appropriate halo. The potential, given the forms (\ref{eq:1Dssvars}), satisfies the same equation above the plane as (\ref{eq:sslaplace}). The boundary condition on a discrete spiral is  difficult to apply with the modal solutions of this equation. We use instead the general self-similar Poisson integral (\ref{eq:Poisson1}). 

We write the surface density in the form (recall the more physical expression of the delta function) 
\be
\sigma=\frac{\delta_D(\xi-\xi_s)}{r}\lambda(r)\equiv \Sigma(\xi)(\delta r),\label{eq:spiralsigma}
\ee
where the linear mass density has the self-similar form $\lambda=\Lambda (\delta r)^2$. Together these last expressions give
\be
\Sigma(\xi)=(\Lambda\delta)\delta_D(\xi-\xi_s). \label{eq:ssspiralsigma}
\ee 
For more than one arm there would be a sum over the various $\xi_s$ (each with its own $K_s$, giving an effective $K(\xi)$) but for present purposes we take only one arm at $\xi_s=0$. It should be clear how to generalize to two or more arms from the following.

The physical potential now follows for $a=0$ from equation (\ref{eq:Poisson1}) as \be
\Phi =-\frac{G\Lambda}{\sqrt{2}}e^{-(1/2)x}\int_{-\infty}^1~\frac{e^{(5/2)x'}dx'}{\sqrt{\cosh{(x'-x)}-\cos{(\xi_s-\phi-(\epsilon/\delta)x')}}}.\label{eq:ssspiralpot}
\ee
We have taken a finite disc of radius $R_{max}=1/\delta$ and $x\equiv \delta R\equiv \ln{\delta r}$. The physical potential is written rather than the scaled potential $\Psi$ since the potential, being long range, might scale differently from the local quantities such as the surface density and the DF. In such a case the self-similar Poisson equation would have to start again from equation (\ref{eq:spiralPoiss}), but since we do not use this equation we can ignore this requirement.

We have taken the physical delta function to evaluate the integral over $\phi'$ at the average value $-(\epsilon/\delta)x'+(\xi_1+\xi_2)/2$. The average $(\overline{\xi}\equiv(\xi_1+\xi_2)/2\equiv \xi_s)$ for simplicity for small $\xi_2-\xi_2$. However this creates a logarithmic singularity due to a rigorous delta function that must be avoided. This can be done formally by keeping $\xi_s$ and $\overline{\xi}$ distinct.  In our present example we set $\xi_s=0$ and discuss only one arm.  For several arms, $\Lambda=\Lambda(\xi_s)$ and there would be a sum of terms over $\xi_s$, which label is manifestly an integral of the motion for the CBE. 

To find the potential on a spiral $\xi=constant$ we constrain $\phi$ in equation (\ref{eq:ssspiralpot}) to equal $\xi-(\epsilon/\delta)x$ This gives finally 
\be
\Phi =-\frac{G\Lambda}{\sqrt{2}}e^{-(1/2)x}\int_{-\infty}^1~\frac{e^{(5/2)x'}dx'}{\sqrt{\cosh{(x'-x)}-\cos{((\epsilon/\delta)(x-x')-\xi)}}}.\label{eq:ssspiralpotF}
\ee
 We observe that there is a logarithmic singularity on the line $\xi=0$ at $x'=x$ for all $x$ since the difference of functions under the square root goes to zero as $x'-x)^2$. However this may be avoided by retaining $\overline{\xi}\ne 0$ in the argument of the cosine. The integral is readily evaluated numerically for $\overline{\xi}$ as small as a milliradian.

\begin{figure}
\begin{tabular}{cc} 
\rotatebox{0}{\scalebox{.4} 
{\includegraphics{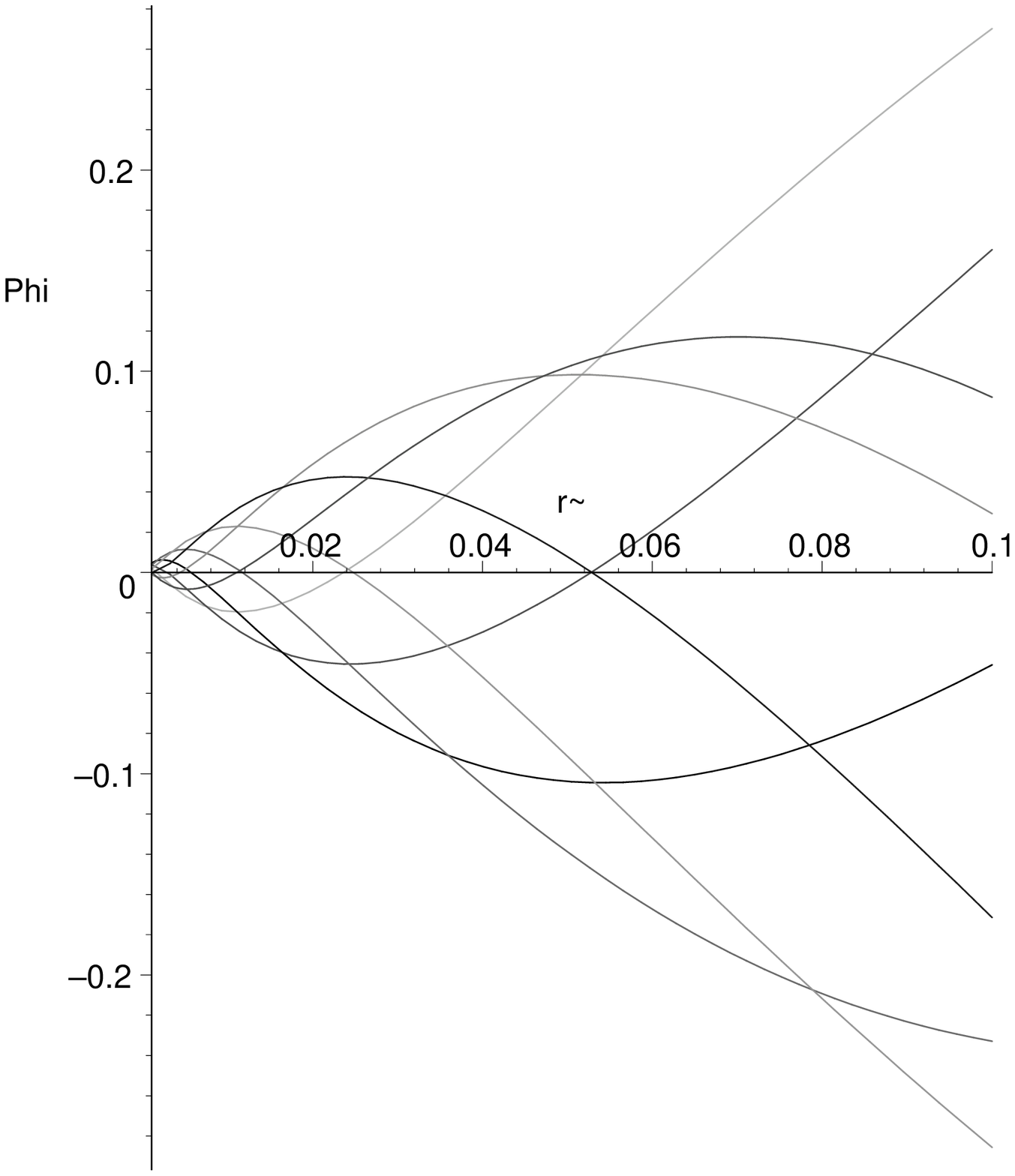}}}
\rotatebox{0}{\scalebox{.4} 
{\includegraphics{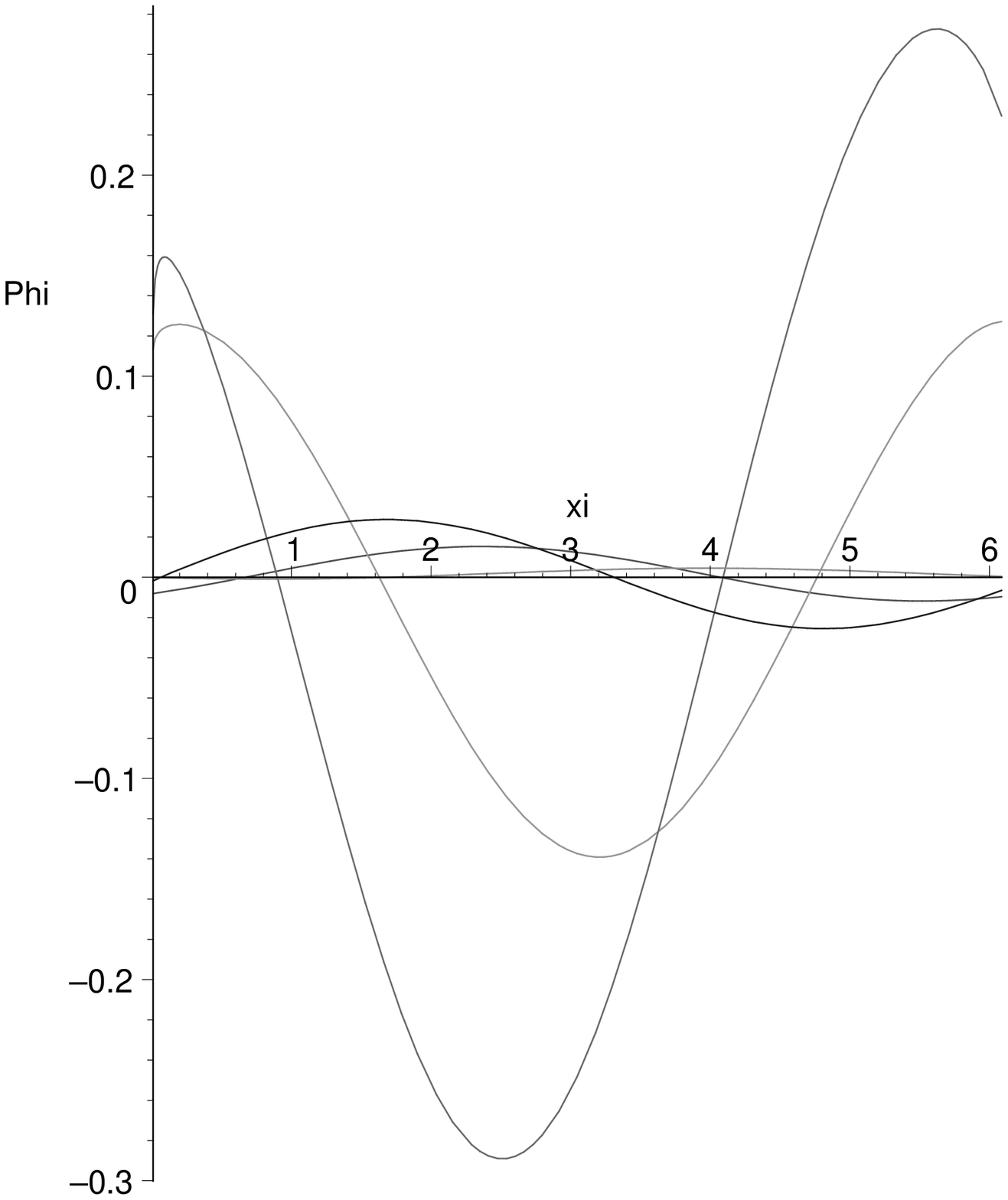}}}
\end{tabular}
\caption{The left hand figure shows the run of the potential in $r$ at small $r$ for different angles. Labelling the curves at $r=+0.1$ and starting from the top the angles are $\{7\pi/4,3\pi/2,\pi/3600,\pi/4,5\pi/4,\pi/2,\pi,3\pi/4\}$ . The right hand figure shows the run of the potential in $\xi$ at different radii. Labelling the curves at $\xi=5$, the radii in units of the disc radius are $\{0.1, 0.05,0.001,0.005,0.01\}$. The constant value of the potential at $r=0.001,\xi=\pi/3600$ equal to $5.226684385$ has been removed in all cases so as to show better the variations.}     
\label{fig:deltapot}
\end{figure}

We have evaluated the integral (\ref{eq:ssspiralpotF}) numerically both as a function of $r$ (recall that $x=\ln{\delta r}$) and of $\xi$. We see the results in figure (\ref{fig:deltapot}).

We conclude that the single armed disc is weakly unstable (outward force where the gradient in $r$ is negative), which has to be corrected by a suitable variation in the halo. More interestingly we see that for $r<0.1$ (in units of the disc radius) $\Phi$ is rather constant to within a few per cent (recall that a constant has been removed from the potential of the order $5.227$). The scaled potential $\Psi$ is certainly not independent of $r$ and must be corrected to the quadratic form.
 In deriving the Boltzmann equation we have assumed  that a  halo maintains the self-similar  harmonic potential. There would only be a weak correction to a uniform density halo given the small variability at small radius, but the necessary `conspiracy' remains.

Physically, we have found in this section a way of constructing discrete rigidly rotating spiral arms that are stabilized by a halo. The self-consistent DF is plausibly given by equation (\ref{eq:1DphysDF}) although we can not study the self-consistency condition without knowing the potential above the plane. The particles move purely along these arms in the rotating frame. There is no net flux of particles along the arms because of the isotropy of the DF, so a particle going out at one stage is returning along the arm after a period in the rotating frame.    
\section{Discussion and Conclusions}

In this study we have looked at spiral structure in  finite thin discs from the perspective of stationary self-similarity in a rigidly rotating frame of reference. It is necessary to appeal to a suitably modified uniform density halo (dark or otherwise) to ensure the existence and the stability of such a structure.

 We have examined the nature of the disequilibrium in a finite disc that has the linear surface density profile. In the case of axi-symmetry we have estimated the perturbation to the uniform density halo required for stable self-similarity. A kind of disc-halo conspiracy is required in general. In later sections we find the form of the potential in the plane of a discrete spiral arm, which also requires corresponding perturbations to the halo.     

We found that the `Kalnajs DF' of equations (\ref{eq:intssP}) and (\ref{eq:intssF}) is the same as the self-similar, uniformly rotating, disc DF for collisionless matter in the rotating reference frame. This is so even though the surface density is not self-similar, although the potential that it inherits from a MacLaurin spheroid is.  We were able to fit this DF into a dimensional sequence wherein it characterizes a two dimensional distribution, while $F\propto |E|^{1/2}$ is the correponding distribution in one dimension and $F\propto |E|^{-3/2}$ is the equivalent distribution in three dimensions. When one relaxes self-similarity one obtains  the slightly modified Jeans' distribution functions in equations (\ref{eq:intF}) and (\ref{eq:intaxiF}). We remarked that the DF (\ref{eq:intaxiF}) is nevertheless reduced to (\ref{eq:intF}) if one supposes that $E/\Omega j_z$ is constant, as is the case for spiral waves.

The equivalent truncated self-similar disc has the same DF and harmonic potential as does the Kalnajs disc, but it has a different upper limit in energy space.  This limit is determined by the self-similarity and is quadratic in radius, being proportional to $\Omega^2r^2$. This is the only difference with the Kalnajs disc, given the existence of a halo that renders the disc potential harmonic. The self-similar disc can not exist in isolation. 

  In this same section we reviewed the scaling symmetry of the Poisson integral for a disc potential and so found the general scaling for finite discs. In the particular case due to Kalnajs \cite{K71}, wherein the potential is expanded in spiral modes, we saw that the similarity class in the sense of \cite{CH91} was $a=5/4$.
   
In section 4 we studied the corrections to a truncated self-similar disc potential that are necessary to obtain a harmonic potential. This is a measure of the disc-halo conspiracy required. Continuing with the self-similar disc including a surface density linear in radius, we derived briefly the potential structure above the plane of the disc. Its main function is to provide a background medium for the non-linear spiral arms discussed in the following sections.

In section 5 we studied first  aperiodic, non-linear, spiral modes. Each of these require spiral discontinuities that should correspond physically to collisionless `shocks', perhaps including the possibility of hydraulic jumps \cite{MC98} . This built-in mechanism for dissipation probably justifies focussing on trailing arms, although formally both cases are possible. The self-consistency condition (\ref{eq:aperselfcon}) requires  discrete spiral arms for constant $K$ and constant fractions of collisionless matter. There is a mode in which the arms  may come in clusters. 

It is interesting to note in this connection that the one-armed structures are constrained to be more open than a winding angle of $45^\circ$. Since the surface density increases outwards, the one-armed structures might resemble more one-sided bars than rings. There is no such restriction for two-armed structures so that they might well be mistaken for rings at radii corresponding to the physical end of the disc.

We turned next in this section to study true periodic spiral perturbations and looked at the individual modes. In this case there are no discontinuities, but other considerations remain much the same as in the aperiodic case. The self-consistency conditions (\ref{eq:selfconperi}) and (\ref{eq:m2modselfcon}) lead once again to discrete arms on using the self-similar DF with $K$ and fractions $f_m$ and $f_a$ constant.   
By using the self-consistency condition for a single mode we have assumed that each mode dominates in turn. We avoid that assumption finally by constructing isolated spiral arms that necessitate only an additional stabilizing halo.
 
To assist this latter construction we changed coordinates from rotating plane polars to rotating `spiral coordinates'. The Boltzmann equation in these coordinates has its uses for discs, but we employed it here to describe a model for a spiral arm in which the motion is wholly parallel to the arm in the rotating frame. The motion is such in the inertial frame that a particle will execute a loop in which both its outward and inward motions  are parallel to the arm. By using an approximate delta function for the surface sensity, the arm may be constructed to have a finite extent in the spiral coordinate. It is thus a bundle of spirals all with the same winding angle. 

Strictly one must avoid concentrating all of the mass on a line since then the arm potential diverges and can not be corrected by a finite halo. 
Accepting this, the potential is found to be only  weakly varying at small radius by evaluating the scaling Poisson integral. Thus the halo may readily impose the harmonic potential without too much distortion. In this case the distribution function on the arm is given by equation (\ref{eq:1DphysDF}). 

Our treatment has not permitted us to say anything about time dependence, except that such structures would not exist in the absence of a suitable halo. The stability is most readily achieved if the halo also is close to the self-similar class of the disc in three dimensions. This is a kind of `disc-halo conspiracy'. Physically these structures can only be taken literally in the central regions of finite discs embedded in halos. However this may not be unreasonable in the central bulge regions of spiral galaxies. The pattern speed of the spirals is one of rigid rotation, but the collisionless particles generally move relative to the arms. This motion is either in and out of the arms in two dimensions or along the arm in the narrow arm limit.

Finally one may well ask why should such an apparently contrived system be of interest in reality? It does appear that uniformly rotating discs exist in the bulges of spiral galaxies, but they might of course be of the Kalnajs type. The only distinguishing feature would be the linearly increasing surface density of the self-similar disc. Moreover it is clear that the halo and the uniformly rotating disc would have had to form in a coordinated fashion. 

The origin of the kind of conspiracy required may well be the coincidence of both the uniform density core and the uniformly rotating disc with the $a=0$ class of self-similarity. The formation of a {\it dominant} core provides the potential necessary for the equilibrium of the self-similar disc. Material falling onto the plane from a growing core would tend to produce a linearly increasing surface density {\it provided} that it also obeys the self-similar symmetry. That is, since there can be no fixed scale, the mass must fall from a height that increases linearly with $r$ with a density that is everywhere constant. Moreover the thickness of the slice $\Delta r$ associated with the mass at $r$, although small, must also increase linearly as  $kr$ for small $k$. Thus as we have often concluded, the self-similar disc must be truncated at a finite radius $R$ such that $kR<<r$.   
\section{Acknowledgements}

This work was supported in part by an operating grant to RNH from the
Canadian Natural Sciences and Engineering Research Council. An anonymous referee offered insightful criticism.

\label{lastpage}

\end{document}